\documentstyle[aps,prb]{revtex} 
\begin {document} 
\draft 
\preprint {MSC Report No.8151} 
\title {Onset of Superfluidity in $^4$He Films Adsorbed on
Disordered Substrates} 
\author {P. A. Crowell,\cite {byline1} F. W. Van Keuls,\cite {byline2} and J. D. Reppy}
\address {Laboratory of Atomic and Solid State Physics and Materials
  Science Center, Clark Hall, Cornell University,\\
  Ithaca, New York 14853-2501}
\date {\today}
\maketitle

\begin {abstract}
We have studied $^4$He films adsorbed in two porous glasses, aerogel and Vycor,
using high precision torsional oscillator and DC calorimetry techniques.    Our investigation focused on the onset of superfluidity at low temperatures 
as the $^4$He coverage is increased.   Torsional oscillator measurements of the $^4$He-aerogel
system were used to determine  the superfluid density  of films with transition temperatures
as low as 20~mK.
Heat capacity measurements of the $^4$He-Vycor system probed the excitation spectrum
of both non-superfluid and superfluid films for temperatures down to 10~mK.  Both sets of
measurements suggest that the critical coverage for the onset of superfluidity corresponds to a mobility edge in the
chemical potential, so that the onset transition is the bosonic analog of a superconductor-insulator transition.     The superfluid density measurements, however, are not in agreement with the scaling theory of an onset transition from a gapless, Bose glass phase
to a superfluid.  The heat capacity measurements show that the non-superfluid phase
is better characterized as an insulator with a gap.   
\end {abstract}


\narrowtext
\twocolumn

\section {Introduction}

For a wide variety of substrates,  adsorbed films of $^4$He thicker than
approximately two atomic layers are superfluid at zero temperature.  
As the temperature is increased, a transition occurs to a non-superfluid phase
at a critical temperature $T_c$.
Although the details of this
transition depend on the topology
of the substrate,\cite {Bishop,KW,Cho,Crowell}  the phase diagrams
for films adsorbed on substrates such as Mylar, porous glasses, 
or packed powders are similar.  In each case, there is a line of
transition temperatures in the density--temperature
plane separating non-superfluid from superfluid films.  At 
$T=0$, this line terminates at a {\it critical coverage} $n_c$, below
which superfluidity does not occur.
Only   
cesium substrates, which are not wetted by $^4$He,\cite {Cesium}
and the atomically ordered substrates graphite and 
molecular H$_2$,\cite {Crowell2,Mochel}
lead to exceptions to this general picture.  

Although the existence of a non-zero critical coverage $n_c$ has been
known since the earliest experiments on unsaturated $^4$He films,
the onset of superfluidity as a function of coverage at $T=0$
has received much less experimental attention than 
the superfluid transition at $T=T_c$.  This
has followed in part from the difficulty in making thermodynamic
measurements on extremely thin $^4$He films, but also from a conviction
that the non-superfluid film of coverage $n_c$ forms an essentially
inert pseudo-substrate for the overlying superfluid film.    
Most aspects of thin-film superfluidity can indeed be understood
without any consideration of the critical coverage $n_c$, and the
$^4$He coverage is often renormalized by taking the effective density
of the film to be $n-n_c$.   For example, the superfluid density
of films with transition temperatures above 200~mK is observed
to be proportional to $n-n_c$ for many substrates.\cite {godknows}  

In contrast to the inert-layer model used for the interpretation of
most experiments, 
there has been considerable theoretical work suggesting that the
onset of superfluidity at $T=0$ in $^4$He films is analagous to 
a metal-insulator transition in Fermi systems.  This approach follows
from a suggestion of Hertz, Fleishman, and Anderson\cite {HFA} and
has been pursued in depth by Fisher 
{\it et al.},\cite {Fisher} who exploited the existence of a 
natural order parameter in Bose systems to develop a scaling theory
for the onset of superfluidity in strongly disordered systems.  
The spirit of this approach is that the onset transition is driven
by the competition between exchange, which favors
superfluidity, and the {\it combined} effects of disorder and the
repulsive He--He interactions, which favor localization.  The localized
phase in the case of a $^4$He film adsorbed on a strongly disordered substrate
is presumed to be a ``Bose glass,'' with a gapless
excitation spectrum and a correspondingly nonzero compressibility.  

This paper presents superfluid density measurements of $^4$He films adsorbed in
aerogel glass and heat capacity measurements of $^4$He films adsorbed in Vycor
for coverages close to the critical coverage $n_c$
and temperatures down to 10~mK.   Our measurements demonstrate
that the picture of a superfluid-insulator transition developed
by Fisher {\it et al.} and others\cite {others,BHworks} is qualitatively
appropriate.  The heat capacity measurements, for example, indicate
that a non-superfluid film is far from inert.  In fact, the density
of states of the non-superfluid film
has a characteristic energy which vanishes as the onset
transition is approached from {\it below} in coverage.   We do
not, however, find good quantitative agreement with the scaling
predictions of Fisher {\it et al.} for the temperature dependence
of the heat capacity and the coverage dependence of the superfluid
density and the superfluid transition temperature.  We
find that the non-superfluid film is best characterized as an insulator
with a gap as opposed to a Bose glass.   The scaling behavior of thin superfluid
films is complicated by both the large energy scale characteristic of physical
adsorption and the existence of the thermodynamic superfluid transition at non-zero temperatures.  The thermodynamic transition appears to be the dominant critical
point, at least over the temperature and coverage range of our study.

\section {Background} 
  
 In Fig.~\ref{fig:generic} we show a generic phase diagram for an unsaturated 
$^4$He-film in
the temperature-density plane.  
The film is superfluid at $T=0$  if the coverage exceeds
a critical coverage $n_c$, which is typically 1.5  to 2 atomic
layers.  (All of the substrates discussed in this paper are disordered and do not support
layer-by-layer growth.  An  ``atomic layer'' thus corresponds to  an average coverage of one layer.)  A phase boundary, corresponding to the superfluid transition
temperature $T_c(n)$, extends upward from the point ($n\!=\!n_c$, $T\!=\!0$).
The film is non-superfluid above this line.
The phase diagram of Fig.~1 is representative
of the aerogel and Vycor systems discussed in this paper as well
as a variety of other disordered substrates.
This paper is devoted to three  
questions regarding this general phase diagram.  First,
how can we better
understand the onset transition at $n=n_c$:  does it have scaling
properties like conventional second-order phase transitions?  Second,
what is the
most appropriate description of non-superfluid films?
Finally, what is our understanding  of the thinnest superfluid films,
within a few percent of the critical coverage $n_c$?

\subsection {Bose Insulators}

One view of the onset transition is that the $^4$He film is solid
at densities below $n_c$ and thus cannot support superfluidity.  This 
viewpoint is equivalent to drawing a vertical line in Fig.~\ref{fig:generic} at
$n=n_c$ and therefore makes a distinction between the ``solid'' part of
the film and the ``liquid'' overlayer, which is superfluid at low
temperature.    Although this approach, often called the 
``inert-layer model,'' is sufficient for interpreting 
most superflow measurements,\cite {godknows}  
it eschews particle statistics by ignoring exchange
between the solid layer and the overlying superfluid film.   In its most
orthodox form, the inert-layer model predicts that the various
thermodynamic properties of the superfluid should be extensive with respect to
$n-n_c$, where $n$ is the total density of the film.    In practice, this does not
appear to be the case.  For example, the superfluid density $\rho_s$ is generally
not linear in $n-n_c$ for coverages very close to the onset of superfluidity.  This
observation has to led to some variations on the inert-layer model, including the
suggestion that the compression of the film as it becomes thicker leads to an increase in the
density of the inert layer.\cite {Crooker}

A more rigorous approach is difficult, and, although
outlined almost twenty years ago,\cite {HFA} has been 
pursued in detail only in the last several years.  It is quite
natural to draw an analogy between the metal-insulator
transition in Fermi systems and the onset transition in a
$^4$He film, which is a Bose system.  In both cases, one
can argue that the onset of metallic (or superfluid) behavior
corresponds to the appearance of extended states spanning
the system above a critical density $n_c$.   Unfortunately,
the theoretical machinery developed for Fermi
systems is not readily adaptable to the Bose case because
of the strong repulsive interactions.
There is no meaningful non-interacting limit, 
such as pure Anderson localization,
for a disordered Bose system.

It is possible, however, to consider a pure but strongly interacting
system and then gradually turn on the disorder.  The simplest representation 
of the interactions is a hard-core repulsion $V$ that prevents two atoms
from occupying the same site.  The effects of quantum exchange are
introduced via a hopping matrix element $J$ between
different sites.  This {\it Bose-Hubbard model} 
has been studied 
for one-dimensional (1D) and two-dimensional (2D) lattice
systems.\cite {BHworks} In both cases, the system is ``superfluid''  
(meaning a many-body extended state exists) 
for any density as long as the exchange energy $J$ is greater than
a critical value $J_c$.  For weak but non-zero exchange, the system
will be superfluid as long as there are sites free to accomodate 
hopping.  This condition fails only at commensurate densities,
at which exactly $n$ atoms occupy each site.  In this case, the
system is localized by the on-site repulsion and is entirely
analogous to a Mott insulator in the fermionic case.  A gap for
particle-hole excitations exists, corresponding to the energy
needed to overcome the on-site repulsion when the next atom is
added to the system. 

Disorder, which can be thought of in the simplest case as a random
fluctuation in the on-site repulsion $V$,\cite {Fisher}
has two effects.  First, it leads to localization at 
{\it incommensurate} densities since
atoms will prefer to occupy sites with the lowest on-site
repulsion.  This can be expected to destroy superfluidity
in the case of sufficiently weak exchange.  In addition,
however, the disorder destroys the Mott insulator at commensurate
densities, since the distribution in $V$ implies that the penalty for double 
occupancy will be reduced at some sites, thus softening the gap.
As the strength of disorder is increased further, the gap closes   
even for small exchange, and the Mott insulator disappears completely.
Although the system is now gapless, it is still
localized (by disorder) for small exchange strengths.  This
gapless Bose insulator is often called a {\it Bose glass}.\cite {HFA,Fisher}

\subsection {Bose Glass to Superfluid Transition}

In the case of porous glasses, the adsorption potential is
strongly disordered on 
the energy scale ($\sim 1$~K) characteristic of the exchange
in a superfluid $^4$He film.  In this case, the onset of superfluidity
within the framework of the Bose-Hubbard model is most likely to occur
from the Bose glass phase. A characteristic  phase diagram in the plane of
exchange strength $J$ and chemical potential $\mu$ for the case 
of strong disorder is shown in Fig.~\ref{fig:glass}. In a real experiment, 
the chemical potential is tuned by changing the density, and one follows
a nearly vertical line in Fig.~\ref{fig:glass} as the $^4$He coverage is increased.
The onset of superfluidity
occurs at the point $\mu=\mu_c$.

The situation sketched in Fig.~\ref{fig:glass} is reminiscent of a 
conventional second order phase transition, with the chemical potential
$\mu$ assuming the role of temperature.  Fisher {\it et al.}\cite {Fisher}
have exploited this analogy to construct a scaling theory of the
Bose glass to superfluid transition, based
on the ansatz that  the correlation length $\xi$ for fluctuations
in the phase of the order parameter
diverges at the onset of superfluidity as
\begin {equation}
\xi \sim \delta^{-\nu},
\end {equation}
where the  reduced chemical potential
$\delta = \mu - \mu_c$ parametrizes the distance to the critical
point.
Since the number density and the phase of the order paramater are conjugate
variables, phase fluctuations are linked to density fluctuations, which propagate at
the speed of sound, introducing a time scale as well as the length scale $\xi$ into
the problem.
Fisher {\it et al.} define the characteristic frequency $\Omega$, which is assumed to
vanish at 
the onset of superfluidity according to the power law
\begin {equation}
\Omega \sim \xi^{-z},
\label{eq:omega}
\end {equation}
where $z$ is the {\it dynamic exponent}.   The existence of the dynamic exponent reflects the quantum mechanical origin of the fluctuations in this problem, as opposed to the thermal fluctuations at an ordinary critical point.  

Although rigorous arguments
are presented by Fisher {\it et al.}, we note here that
conventional hyperscaling will hold if we assume that the
correlation volume has a time dimension $(1/\Omega)$ as well as a
spatial part $\zeta^d$, where $d$ is the ordinary spatial dimensionality.  This allows us to write down
an asymptotic form for the singular part $f_s$ of the free energy density at the
onset transition:
\begin {equation}
f_s(T=0) \sim \delta^{(d+z)\nu}.
\label {eq:fs}
\end {equation}
An asymptotic form for the {\it superfluid density} $\rho_s$ on the
superfluid side of the transition follows from finite size
scaling:\cite {FBJ}
\begin {equation}
\rho_s(T=0) \sim \delta^\zeta,\hspace {36pt} \zeta = (d+z-2)\nu.
\label {eq:rhos}
\end {equation}
The singular part of the compressibility,
\begin {equation}
\kappa_s = \frac{-\partial^2 f}{\partial \mu^2} \sim \delta^{\nu(d+z)-2},
\label {eq:kappas}
\end {equation}
is analogous to the singular part of the specific heat
at a conventional phase transition.

In order to make experimental predictions from these scaling laws,
one needs to know the dynamic exponent $z$.  Fisher {\it et al.}
prove that $z=d$ for the case of a Bose glass to superfluid transition.
The proof depends critically
on the Bose glass having a non-zero compressibility, or, equivalently, a gapless
low-energy excitation spectrum.

The theory up to this point makes
predictions for the scaling properties of the superfluid density
only at $T=0$.  Fisher {\it et al.} have extended
their theory to thermodynamic properties at non-zero temperature using
finite size scaling in $1/T$.  In implementing this argument, they
assume that the only energy scale in the problem is set by the
characteristic frequency $\Omega$.  Temperature will thus appear in
the asymptotic free energy in the form $k_B T/\hbar \Omega$.  
Taking $k_B = \hbar = 1$, the scaling-form for the free energy becomes
\begin {equation}
f_s(T) \sim \delta ^{\nu(d+z)} \tilde{f}(T/\Omega).
\label {eq:fst}
\end {equation}
The superfluid transition (at non-zero temperature) occurs at
a universal value $u_c$ of the argument $T/\Omega$, so that
\begin {equation}
T_c = u_c\Omega \sim \delta^{z\nu}.
\label {eq:Tc}
\end {equation}

In order to reduce any of the scaling laws into forms testable by
experiment, we need to relate $\delta = \mu - \mu_c$ to the
density $n$, which can be tuned in the laboratory.  Assuming
$z=d$ and using the inequality $\nu \geq 2/d$ for
the correlation length exponent in disordered systems,\cite {Chayes} 
it can be shown that 
\begin {equation}
n - n_c \sim \mu - \mu_c,
\label {eq:nmu}
\end {equation}
which means that $\delta$ can be replaced by $n-n_c$ in all of
the above scaling laws.  After doing so,
we obtain the scaling laws of Fisher
{\it et al.}:\cite {Fisher}
\begin {eqnarray}
\rho_s(0)  \sim  (n-n_c)^\zeta,  & \zeta \equiv \nu(d+z-2);\label{eq:rhosvsn}\\
T_c   \sim  (n-n_c)^w, & w \equiv z\nu; \label{eq:tcvsn}\\
T_c  \sim  [\rho_s(0)]^x, & x \equiv z/(d+z-2),\label{eq:tcvsrhos}
\end {eqnarray}
where $\rho_s(0)$ is the superfluid density
at $T=0$.  The torsional oscillator experiment discussed below addresses
the validity of these scaling laws for $^4$He films adsorbed in aerogel glass.

\subsubsection {Heat Capacity}

We now consider the low-temperature heat capacity on both sides of the onset
transition.  Technically, the
only requirement for the Bose glass is that the excitation spectrum
be gapless. This admits almost any power-law dependence of $C(T)$,
although for a true glass, {\it i.e.} constant density of states,
$C(T) \propto T$.  On the superfluid side of the onset transition,
the dominant low-energy excitations are long-wavelength third sound
modes, leading to a Debye specific heat, $C(T) \sim T^d$ for
$T \ll T_c$.

At onset ($n=n_c$), the specific heat can be inferred from the scaling
form for
the free energy (Eq.~\ref{eq:fst}).  In order for $f_s(T)$ to remain
non-zero and finite at the transition, the temperature dependent
part $\tilde{f}(T/\Omega)$ must assume the form
\begin {equation}
\tilde {f} (T/\Omega) = \tilde {f}(T\delta^{-z\nu})
\sim (T\delta^{-z\nu})^{\frac{d+z}{z}},
\end {equation}
since the sum of all the powers of $\delta$ in Eq.~\ref{eq:fst}
must equal zero.  At onset,
\begin {equation}
C(T) =T\frac{\partial^2 f}{\partial T^2} \sim T^{\frac {d}{z}}, \hspace {36pt}
(\delta = 0, T\rightarrow 0). \label{eq:catonset}
\end {equation}
Note that if $z=d$, as is believed to be the case for  a Bose glass to
superfluid transition, $lim_{T \rightarrow 0} C(T) \sim T$ at onset.
The temperature dependence at onset  is thus the same as in the Bose glass phase.

\subsubsection {The Inert-Layer Model}

Fisher {\it et al.}\cite {Fisher} demonstrate that in the mean-field
limit, their scaling theory reduces to the inert-layer
model, in which the superfluid film is taken to coexist with a
solid-like $^4$He underlayer.  In this case, exchange between the  
superfluid film and the localized layer is assumed to be negligible,
and the thermodynamics of the system near the onset of superfluidity
can be deduced from consideration of an ideal Bose gas coexisting with
a solid $^4$He pseudo-substrate.

The superfluid density in this limit should be proportional
to $n-n_c$, where the critical coverage $n_c$ is the density of the
solid layer. Thus, $\rho_s(0) \sim (n-n_c)^\zeta$, where 
$\zeta = 1$ for both $d=2$ and $d=3$.   Two more exponents can be
inferred directly from the density dependence of the Bose-Einstein
condensation temperature:
\begin {equation}
k_BT_c = \frac {2\pi\hbar^2}{m}\left[ \frac {N}{V\zeta(d/2)}
\right ]^\frac{2}{d},
\end {equation}
where $V$ is the volume (or area in two dimensions) of the system.
Strictly, $T_c = 0$ for $d=2$ because of the logarithmic divergence
of the zeta function $\zeta (1)$.  This divergence can be removed
through the introduction of a short-length cutoff which accounts for
the finite thickness of the film.  We thus find 
$T_c \sim (n-n_c)^w$ and $T_c \sim [\rho_s(0)]^x$, where 
$w = x = 1$ for $d=2$, and $w=x=2/3$ for $d=3$.

We now propose low-temperature forms of the heat capacity for
the inert-layer model.  For $n > n_c$, the heat capacity of the
{\it ideal} Bose gas has the limiting behavior $C \sim T^\frac{d}{2}$ as
$T \rightarrow 0$.  (The presence of superfluidity requires a Debye
contribution $C \propto T^d$, but this vanishes more rapidly than the
Bose gas contribution in the limit $T\rightarrow 0$.)
Since the inert-layer model
makes no assumptions about the non-superfluid film, it admits
essentially any form of the heat capacity for densities $n < n_c$.

\subsubsection {Summary of the Scaling Laws}

In Table~\ref{tab:scaling}, we summarize the results of the scaling
theory for both the Bose glass and inert-layer models.  
Four possible cases are considered, corresponding to the Bose glass
or inert-layer models in either 2 or 3 dimensions.  In calculating
the exponents from Eqs.~\ref{eq:rhosvsn}-~\ref{eq:tcvsrhos} and~\ref{eq:catonset}, we have used the result
$z=d$ as well as the correlation length inequality $\nu \geq 2/d$.\cite {Chayes}

\section {Experimental Details}

Two experiments will be discussed in this paper.  The first
is a systematic study of the superfluid density of
$^4$He films adsorbed in 91\% porosity aerogel glass for temperatures
between 10~mK and 200~mK.  The second experiment is 
a high-resolution heat-capacity study of $^4$He films adsorbed
in Vycor glass for temperatures from 10 to
200~mK.  The heat capacity measurements were conducted for
coverages on both sides of the onset transition.  Brief accounts
of these experiments have been published previously.\cite{PennState,LT20,VycorPRL}

\subsection {Substrates}

Vycor and aerogel are both porous glasses.  Vycor\cite {Corning}
is prepared by leaching out one phase of a borosilicate glass after
spinodal decomposition.  The resulting matrix has a 
distribution of pore diameters with a maximum at approximately
70~\AA\ and a half-width of approximately 20~\AA.  
Aerogel\cite {aerogel} is prepared by kinetic aggregation of 
silica particles in a gel.  After hypercritical drying to remove the solvent,
the resulting substrate consists of strands of silica approximately
50-100~\AA\ in diameter.  Unlike Vycor, there is no characteristic
pore size, but the structure is correlated on length scales
up to 600~\AA\ for the sample studied in this work.\cite {Crowell}  In spite
of the structural differences, the superfluid densities of thin
$^4$He films adsorbed in Vycor and 91\%\ porosity aerogel near
$T_c$ are
remarkably similar, although the heat capacity singularity at
the superfluid transition seen in Vycor does not appear in 
aerogel.\cite {Crowell} As discussed in Ref.~\onlinecite {Crowell},
the correlated disorder of the substrate is apparently irrelevant for very
thin films of superfluid $^4$He adsorbed on either substrate, since the superfluid correlation
length $\xi$ exceeds all structural length scales.

\subsection {Superfluid Density Measurements}

The superfluid density of $^4$He films adsorbed in 91\%\ porosity
aerogel was measured using the torsional oscillator
technique.\cite {Wong}  The experimental cell was the same as that
described in Ref.~\onlinecite{Crowell}, which discusses
the experimental aspects of the superfluid density measurements in detail. 
Several important parameters are listed in Table~\ref{tab:cell}.  
All of the torsional oscillator data in this paper will be presented in terms
of the {\it superfluid period shift} $\Delta P$ defined below.  These data can be converted
to a superfluid mass by dividing by the {\it superfluid mass sensitivity} of the cell,
0.022~nsec/$\mu$g.      

  The measurements
under discussion here can be divided into two classes.  The first
comprises temperature sweeps conducted in a manner 
identical to that described in Ref.~\onlinecite{Crowell}.  For each   
$^4$He coverage, the resonant period $P$ and the amplitude
of the torsional oscillator were measured as a function of 
temperature.  To determine the superfluid period shift $\Delta P$,
we first subtracted the temperature-dependent background of the empty cell
from the raw data.  The data above the superfluid transition
temperature $T_c$ were fitted to a constant.  The period
data were then subtracted from this baseline, giving the
superfluid period shift $\Delta P(T)$.  The period shift at $T=0$,
$\Delta P(0)$, was determined by fitting $\Delta P(T)$ to a 
horizontal line at the lowest temperatures.  Uncertainties due
to the background subtraction and the fact that  our lowest  reliable
measurements could be made only above 10~mK limited the precision
of our $\Delta P(0)$ measurements to about $\pm 0.1$~nsec.

The superfluid transition temperature $T_c$ was determined with a typical
uncertainty of $\pm 0.2$~mK  by locating the break in $\Delta P(T)$
at the transition.   The most precise way to do this was to
drift slowly ($dT/dt \sim 0.2-0.3$~mK/hr) through the transition
while recording the resonant period and amplitude of the torsional oscillator. 
 The result of such
a drift measurement for $n=35.7$~$\mu$moles/m$^2$  is shown in 
Fig.~\ref{fig:drift}, for which $T_c = 38.6\pm0.2$~mK.  These
data also show two dips in the amplitude.  The large feature,
near 38~mK, is certainly a sound resonance since the 
period, which is measured using a phase-locked loop,
shows an s-shaped distortion.  The smaller peak, which seems
to coincide with the superfluid transition, may be due to
a cascade of resonances.  We do not know why the resonance
associated with the large peak is so prominent, since a simple
calculation shows that it does not correspond to a low-order
third sound resonance.  For the thinnest films, the resonance
could be seen even when $\Delta P(T)$ was almost obscured
by noise and uncertainties in the background subtraction.  

For films with transition temperatures above 50~mK, we could also determine
$T_c$ by fitting the period shift data near the transition to a
power law $\Delta P(T) = At^\zeta$, where $t=1-T/T_c$.\cite {Crowell}
The $T_c$ determined in this fashion was typically 0.2-0.3~mK
below the temperature at which $\Delta P(T)$ was non-zero
within the resolution of our measurements. In this sense, the
superfluid transition for films on aerogel is considerably
sharper than that observed in {\it rods} of Vycor,\cite {Bishop,Crooker} where
a ``foot'' in $\rho_s$ extends 1-2~mK above the superfluid $T_c$
as determined from power-law fits.\cite {annealing}   

\subsection {Heat Capacity Measurements}

The calorimeter, shown
in Fig.~\ref{fig:calorimeter}, was constructed from thin-walled silver.
Fourteen disks of thickness 0.3-0.6~mm 
were cut from a slab of Vycor by first coring the slab with a
slurry cutter and then slicing off the disks using a diamond-impregnated wire
saw.  Cutting from a slab with the leaching plane parallel to the surface produces
a more homogenous sample than is obtained from rods of Vycor, in which the
boron-rich phase is leached from
the perimeter.\cite {Robinson}  After cutting, the disks were cleaned in hydrogen
peroxide and then baked in vacuum.  
 A silver coating, $\sim$0.5~$\mu$m thick, was then evaporated on both
sides of each disk.  The edges were left uncoated to allow the $^4$He
to penetrate the sample.   Each disk was then glued to two 0.025~mm thick Ag
foils, which had previously been diffusion-welded to the cap of the calorimeter.
The calorimeter was sealed with Emerson Cumming Stycast 1266 epoxy.
A  Pt-W wire heater was wound around the body of the calorimeter, and a 1~k$\Omega$
RuO$_2$ resistor was glued to the cap of the cell.  The calorimeter was supported by
four Vespel posts and was attached to the low-temperature stage
of a PrNi$_5$ demagnetization cryostat.  The thermal contact to the stage was
controlled using a tin-wire superconducting heat switch. 

 A low thermal mass DC magnetization thermometer, shown in Fig.~\ref{fig:thermometer}, was essential for measurements below 20~mK. The actual
sensor in this thermometer is a miniaturized version of the design of Greywall and
Busch.\cite {GB}   2.5~mgrams of cerous magnesium nitrate (CMN) powder, consisting of 
grains less than 75 microns in diameter, was mixed with an equal volume of 2-4 micron
silver powder and then pressed between eleven 0.025~mm thick silver foils.  Before pressing
with the powder, the foils were
annealed, perforated to enhance the adhesion of the Ag-CMN powder, and their ends were 
diffusion welded together.  After pressing, the  assembly was potted in Stycast 1266 epoxy and machined into a cylinder 3~mm in diameter.  A 2 $\mu$H NbTi wire coil, which was part of a superconducting flux transformer, was then wound around the cylinder,  and the foils 
were attached to a Ag wire using a crimp joint.  This assembly was then repotted in epoxy
and glued to a Vespel post, which was attached to the low temperature stage of the refrigerator.   Two Nb shields were used to trap a small (25-50 ~Gauss) magnetic field and
 for magnetic shielding.  The wires for the flux transformer were twisted and then threaded through a Nb capillary to a junction box on the still  of the dilution refrigerator.
The capillary was filled with silicon oil in order to prevent any vibrations of the twisted pair.  The remainder of the flux transformer passed through a feedthrough into the bath space and was attached to the input coil of a commercial DC SQUID.\cite {QD}

The operation of the thermometer is identical in principle to the high-resolution thermometer developed for $\lambda$-point experiments by Lipa and Chui.\cite {Lipa}
Some details have been published previously by our group.\cite {Wong}  The magnetization is measured by monitoring the current in the 
flux transformer, which changes so as to keep the total flux enclosed 
by the superconducting loop constant.  The sensitivity of the thermometer is determined
by the applied field.
The price paid for the sensitivity of the DC magnetization thermometer is the need to count flux,
since the dynamic range of the flux-locked loop is limited, and so the loop must therefore  be reset
periodically.  In practice, we regularly exceeded the slew rate of the SQUID electronics and the flux count therefore became meaningless.  This inconvenience was ameliorated in our case by the availability of the heat
switch on the calorimeter, which could be closed to check the magnetic thermometer against a $^3$He melting curve thermometer on the stage at any point during the
experiment.    

The heat capacity measurements were conducted using the adiabatic calorimetry technique.
A current pulse was applied to the heater, and the resulting thermal response of the cell
was recorded.    Two response curves, recorded at temperatures of 102.1 and 18.6~mK for
a coverage of 27.1 $\mu$moles/m$^2$, are shown in Fig.~\ref{fig:decay}.    These data are representative of
two classes of response curves  that we found.  Above 80~mK, the temperature of the cell decayed
exponentially after the pulse, and the temperature step $\Delta T$ was determined simply
by extrapolating the exponential decay back to the midpoint of the heat pulse.  At lower
temperatures, the decay became non-exponential, as can be seen in the lower panel
of Fig.~\ref{fig:decay}.  Since we could fit the long-time part of the decay to an exponential,  we first believed that the short-time behavior reflected a thermal overshoot
due to poor thermal contact between the $^4$He film and the calorimeter.  If we extrapolated the long-time behavior back to the midpoint of the pulse, however, the inferred
heat capacities were clearly too high.  This suggested that either heat was leaking out
of the calorimeter on short time scales, hence leading to an overestimate of the heat capacity, or that the anomalous decay was due to some internal relaxation mechanism, perhaps reflecting a significant heat capacity in poor thermal contact with the rest of the
calorimeter.
The former possibility was unlikely since the superconducting heat switch provides
its best thermal isolation at the lowest temperatures.  We suspected that the Vycor itself
might harbor a thermal reservoir and modeled the thermal relaxation inside the cell assuming the existence of a reservoir that was weakly linked to the $^4$He film.\cite {LT20}.    This provided a reasonable fit to the decays, but we found that the thermal conductivity to the weakly linked reservoir depended on both coverage and temperature.
A similar coverage-dependent coupling between the $^4$He film and the Vycor itself
has been observed in ultrasound measurements.\cite {Mulders}
We decided to proceed on the assumption that the $^4$He film
equilibrated rapidly with the thermometers, in which case  the temperature step $\Delta T$ 
for determining the specific heat should be extracted from the response of the calorimeter
at short time scales.  We tested this hypothesis by dosing approximately 2 monolayers of $^3$He into the calorimeter at the end of the run.  This film had a heat capacity approximately
60 times larger than any of the $^4$He films in our
experiment.  The
$^3$He film equilibrated within 120 sec after each heat pulse,  justifying our assumption
about the short time-scale behavior of the calorimeter at low temperatures.  

The heat capacity for a coverage of 27.6 ~$\mu$moles/m$^2$ is shown in
Fig.~\ref{fig:cpexample} after dividing by the temperature $T$.     These data,
taken with the magnetization thermometer, show a cusp at 20~mK, which is a signature of the superfluid transition.\cite {Finotello,Murphy}
 (Since we did not take a BET isotherm for this sample, the surface area
was determined by comparing our measured superfluid transition temperatures as a function of $^4$He dosage with those of Crooker {\it et al.}\cite {Crooker})   The 
ordinary low-temperature limit of our measurements was 10~mK.  This was determined  by the
internal heat leak of the calorimeter, which decayed from approximately
0.8 to 0.4~nW over the course of three months.  This  large heat leak,
which we suspect was due to the Vycor in the calorimeter, required us to run with the
heat switch partially open at the lowest temperatures.   Obtaining a proper  tuning of the
heat switch was very tedious, and this turned out to be the primary impediment to using
the calorimeter below 10 mK.

\section {Results}

\subsection {Superfluid Density}

In Fig.~\ref{fig:alldat}, we show the period shift data for twelve
coverages of $^4$He adsorbed on aerogel.  
The superfluid period shift at $T=0$,
$\Delta P(0)$, was obtained by fitting the low-temperature part of 
each curve to a horizontal line.  The transition temperature $T_c$
was determined for each coverage using the drift technique described
above.  In Fig.~\ref{fig:deltap0}, we show
$T_c$ and $\Delta P(0)$ as a function of coverage.  The curves show
fits to power laws
\begin {equation}
T_c(n) = T_0(n/n_c - 1)^w
\label{eq:tcpowerlaw}
\end {equation}
and
\begin {equation}
\Delta P(0) = D_0(n/n_c - 1)^\zeta,
\label{eq:dp0powerlaw}
\end {equation}
where we find $w =  1.59 \pm 0.06$ and $\zeta = 1.64 \pm 0.04$.
The critical coverage $n_c = 33.89 \pm 0.08$  $\mu$moles/m$^2$ was determined
from the fit of $T_c$ versus coverage and then held fixed for the fit of
the period shift data.
The final relation of
interest is the dependence of $T_c$ on $\Delta P(0)$, which is
shown in Fig.~\ref{fig:tcvsdeltap0}.  In this case, we have fitted 
the data to the power law 
\begin {equation}
T_c = A[\Delta P(0)]^x,
\label{eq:tcp0powerlaw}
\end {equation}
where we find $x = 0.95 \pm 0.02$.  The measured exponents are listed in Table~\ref{tab:exponents2} along with corresponding values we obtain from
fits of the data of Crooker\cite {Crookerthesis} for the $^4$He-Vycor system.

We will discuss the exponents of Table~\ref{tab:exponents2} in more
detail below.  Another point of interest is the size of
the ``critical region'' for the onset transition.  Since the
fits of the various power laws are poor (by the standards generally applied to
critical phenomena), 
the usual practice of looking
at deviations from the power-law behavior is not particularly instructive.
Another approach is simply to see how the data near onset compare
with those further away from the transition.  We do this for $\Delta P(0)$
as a function of coverage in Fig.~\ref{fig:dp0all}.
The highest coverage in this figure has a $T_c$
of approximately 1~K.  A significant departure from linear behavior is
seen only for films with $\Delta P(0) < 1.5$~nsec, or $T_c < 150$~mK.
The linear dependence of $\Delta P(0)$ on $n$ at higher coverages is
consistent with the inert-layer model of the onset transition, so
the data of Fig.~\ref{fig:dp0all} indicate only a small coverage
regime in which fluctuations associated with the onset transition might be
relevant.
The evolution of $T_c$
with $\Delta P(0)$ is even more striking in this regard, since,
as shown in Fig.~\ref{fig:tcall}, the nearly linear behavior observed for
thin films continues for films with $T_c$'s up to 0.7~K.  [This linear dependence
is suggestive of the Kosterlitz-Thouless-Nelson (KTN) relation for a two-dimensional
superfluid,\cite {KTN}  for which $\rho_s(T_c^-)/T_c = 8.73$~$\mu$moles~m$^{-2}$K${-1}$,
where $\rho_s(T_c^-)$  is the limiting superfluid density as the transition temperature
is approached from below.     
For reasons of comparison in Fig.~\ref{fig:tcall}, we have converted the period shift
to {\it areal superfluid density} using the calibration $\rho(s)/\Delta P = 1.24$~ $\mu$moles~m$^{-2}$nsec$^{-1}$.
The slope of the KTN line, which is drawn dashed in the figure,  is about twice that of a linear fit to the
data of Fig.~\ref{fig:tcall}.   Although we expect $\rho_s(T_c^-)$ to be reduced from
$\rho_s(0)$ by elementary excitations and vortex pair screening, the discrepancy between
the two slopes is not resolved in a full fit   
of the superfluid density data to the finite size theory \cite {MG}  for a superfluid
film adsorbed on a cylindrical strand. The best-fit coverage for each data set that is about half the actual coverage. \cite {Crowellunpublished}    We do not believe, however,  
that the failure of this naive application  of the KTN relation says anything about the effective dimensionality of the film, which will be discussed further below.]

\subsection {Heat Capacity}

The $^4$He-Vycor heat capacity data, divided by temperature, are shown for ten coverages in Fig.~\ref{fig:alldat2}.   Each of the data sets in the lower panel of the figure shows
a sharp cusp, corresponding to a superfluid transition.\cite {Finotello}   The superfluid transition
temperatures  determined from the locations of the cusps are shown as closed
circles in Fig.~\ref{fig:tbtc}.
The critical coverage $n_c$ for the onset of superfluidity
is determined by extrapolating the $T_c$ data to zero temperature.
For temperatures $T < T_c$, the heat capacity of the superfluid
films is roughly quadratic in temperature, while it depends nearly linearly on temperature for
$T>T_c$, with a small quadratic correction that increases with coverage.  In the vicinity of the superfluid transition, there is a small peak in $C/T$
due to critical fluctuations.\cite {Finotello,Murphy}  The size of the peak decreases with
decreasing thickness.  As can be seen in Fig.~\ref{fig:cpexample}, this peak can barely 
be resolved for the thinnest superfluid film ($T_c$ = 20.5~mK).  

   At high temperatures, the heat capacity of non-superfluid films, shown divided by temperature in the upper panel of Fig.~\ref{fig:alldat2}, depends linearly on temperature,
but C/T drops rapidly at low temperatures.  We define a crossover temperature $T_B$
separating these two regimes to be the point at which $C/T$ is half of its high-temperature value.  $T_B$ is shown as a function of coverage in
Fig.~\ref{fig:tbtc}, along with the $T_c$'s measured for the superfluid
films.    This plot shows clearly that $T_B$ vanishes as the onset transition is
approached from below.    (Although our definition of $T_B$ is somewhat
arbitrary, any reasonable choice of a characteristic temperature for the
crossover would show the same coverage dependence.)  For  $T>2T_B$, there is not
much qualitative difference between the heat capacity of a non-superfluid film
and that of a superfluid film for $T>T_c$.  This is emphasized in the
three-dimensional representation of our data shown in Fig.~\ref{fig:3D}, in
which there is a wedge-shaped plateau in the vicinity of  the onset coverage
$n_c = 27$~$\mu$moles/m$^2$. Above 20~mK, the heat capacity isotherms, which are
cross-sections of the 3D surface, show no feature at $n_c$.  The peaks in the
isotherms become progressively narrower in coverage as the temperature is
reduced, but we cannot reliably extrapolate our data to $T=0$.

\section {Discussion}

The data introduced above indicate that the concept of a superfluid-insulator transition in
$^4$He films is qualitatively appropriate.    Both sets of measurements indicate 
the existence of a critical coverage $n_c$ at which the superfluid  transition temperature $T_c$ vanishes.   The
heat capacity data are particularly instructive in this regard, since there is a characteristic
energy $k_BT_B$ for nonsuperfluid films which also vanishes as $n_c$ is
approached from {\it below} in coverage.

\subsection {Scaling Theory}

\subsubsection {$\rho_s$ measurements}

As can be seen by comparing the exponents of Table~\ref{tab:exponents2}
with the predictions of Table~\ref{tab:scaling}, the experimentally
determined exponents for the onset transition  are not consistent with the scaling theory for either $d=2$ or $d=3$.    
We have also seen in Figs.~\ref{fig:dp0all} and~\ref{fig:tcall} that
the departure from the behavior seen at higher coverages ($T_c > 200$~mK)
is actually quite small.  Some curvature is apparent in $\Delta P(0)$
at the lowest coverages, but the relation between $T_c$ and 
$\Delta P(0)$ remains nearly linear for $T_c$'s from 700~mK down to the
lowest $T_c$ in our study (20~mK).  These observations lead us to
question whether our measurements are in the asymptotic regime of reduced
coverage $n-n_c$ in which the scaling theory is valid. To address this 
question, we return to the asymptotic form of the free energy,
Eq.~\ref{eq:fst}.  This scaling form comprises the product of the free energy
at $T=0$ and a function of $T/\Omega = u_cT/T_c$.  This implies
\begin {equation}
\frac {\rho_s(T)}{\rho_s(0)} \sim \tilde u (u_c T/T_c)
\end {equation}
in the asymptotic regime, where $\tilde u$ is a universal function.  Thus, upon scaling the
period shift $\Delta P$ by $\Delta P(0)$ and the temperature by
$T_c$, all of the $\Delta P(T)$ data in the asymptotic regime
should collapse on to a
universal curve.  
As can be seen in Fig.~\ref{fig:bgscale}, our data do not scale in this 
fashion.  In fact, the scaled data in this figure 
drift continuously toward the origin
as the coverage is reduced.  There is no indication that the data 
are even approaching a universal curve.  The scaling plot demonstrates
explicitly that the properties of superfluid films at non-zero temperature
are {\it not} determined by the single energy scale $\Omega$.  Therefore,
we should not expect the scaling laws (Eq.~\ref{eq:tcvsn} and~\ref{eq:tcvsrhos})
for $T_c$ as a function of $n$ or $\rho_s(0)$ to hold in the coverage range of
our experiment.

Although we have shown explicitly that the scaling theory does not hold
at non-zero temperature ({\it i.e.} $T > 20$~mK), we must still consider the possibility
that it holds at $T=0$.   Only one of the scaling laws,  
$\rho_s(0) \sim (n-n_c)^\zeta$, applies in this case, but the experimental
result $\zeta = 1.64\pm 0.04$ is well below the theoretical bounds
$\zeta \geq 8/3$ for $d=3$ or $\zeta \geq 2$ for $d=2$.   Since,
as shown in Fig.~\ref{fig:dp0all}, the data approach the inert-layer
prediction at higher coverages, it is reasonable to ask if our
low-coverage data fall in a crossover regime in which quantum 
fluctuations are relevant but not yet dominant.  This is certainly
possible, but it leads one to ask why the critical 
region in coverage should be so small.  In order to address this question, we convert the coverage to the chemical potential $\mu$ using the
Van der Waals relation
\begin {equation}
\mu(n) = - \frac {\alpha}{n^3},
\label {eq:VDW}
\end {equation}
where $\alpha = 27$~K (layer)$^3$ for $^4$He on glass.\cite {alphareference}  Although  the chemical potential  is difficult to measure in this coverage range,  third sound measurements, which determine $\mu(n)$ indirectly,  are consistent with the form of Eq.~\ref{eq:VDW}.\cite {Bishop}  Assuming a monolayer
coverage of 13 $\mu$moles/m$^2$, $n_c \approx$~2.6 layers for
$^4$He-aerogel and $n \approx$~3.0~layers for a film with a $T_c$
of 200~mK.  The chemical potential thus shifts from about -1.5~K
at onset to approximately -1.0~K for films with $T_c$'s near 200~mK.   
Given that the characteristic exchange energy for the $^4$He films is on the
order of 1--2~K,\cite {Ceperley}
a chemical potential shift of 0.5~K cannot be considered small.

The
rapid change of $\mu$ with coverage also leads to a technical 
problem in applying the scaling theory:  
the assumption $n-n_c \sim \mu - \mu_c$ is not
correct, even in the narrow coverage range of our experiment.  Assuming
that $\mu$ follows the form of Eq.~\ref{eq:VDW}, we have converted 
coverage to chemical potential and have fitted our data to the power law
\begin {equation}
\Delta P(0) \sim (\mu - \mu_c)^\zeta.
\end {equation}
In effect, this allows us to eliminate the assumptions implicit in
Eq.~\ref{eq:nmu}, with the caveat that we have not
actually measured the chemical potential.  We find an exponent
$\zeta = 2.2\pm 0.1$ by this approach.  The case for the
Bose glass is thus improved in $d=2$, although
the inequality of Eq.~\ref{eq:rhosvsn}  is still not satisfied for $d=3$.   
We note, however, that the theory cannot be applied self-consistently
in $d=2$, since it assumes a diverging correlation length which will
eventually exceed the strand size of the aerogel.  In this case, 
topological arguments lead us to expect that the onset transition 
must be three-dimensional in character, although we cannot exclude
the possibility of an observable crossover regime between 2D and 3D
behavior.   

We thus find that two
conditions of the scaling theory do not apply in the case of
$^4$He adsorbed in aerogel in the coverage range of our study.  
First, there is no single energy scale
that determines the thermodynamic properties at finite temperature.
Second, the chemical potential changes rapidly with coverage,
which makes the asymptotic regime difficult to achieve and 
also implies that the reduced density $n-n_c$ is not simply
proportional to $\mu- \mu_c$.  We recall at this point the
third (and most fundamental) assumption of the theory, which
is that the non-superfluid phase is compressible.  (This assumption is
critical to establishing the relation $z=d$ for the dynamic exponent.)
Since the compressibility is given by $\partial n/\partial \mu$,
Eq.~\ref{eq:VDW} suggests that although $\kappa$ will be non-zero,
it will be considerably lower than for an idealized
$^4$He film such as that considered in the Bose-Hubbard model.\cite {Fisher,others,BHworks}
In effect, the adsorption potential adds a third energy scale to the problem in addition
to the exchange energy and the interparticle repulsion.    The strong binding to
the substrate makes the film much less compressible than it would be if it were
``free-standing."\cite {Zimanyi}  
Even if $\kappa$ is technically non-zero,
the film may be sufficiently incompressible that the softening due to 
quantum fluctuations is negligible.

\subsubsection {C(T) measurements}
 
The heat capacity measurements allow us to test the predictions of the
scaling theory below the onset of superfluidity.  As shown in 
Fig.~\ref{fig:alldat2}, the salient feature of $C(T)$ for coverages below
onset is a characteristic temperature
$T_B$ below which $C/T$ drops rapidly to zero and above which it
approaches a constant.  As shown in Fig.~\ref{fig:tbtc}, $T_B$ vanishes
as the onset of superfluidity is approached from below.
At the onset transition, $C(T)$ is approximately linear over the entire temperature
range, $T>10$~mK, covered by our experiment.

The fact that $C(T)$ is linear in $T$ for $n=n_c$ is in apparent agreement with the
scaling theory of Fisher {\it et al.}\cite {Fisher} (see Table~\ref{tab:scaling}).
Below onset, however,
the data show only a single region of linear
behavior, although the Bose glass model  predicts two regions: one at low
temperature
due to the Bose glass and a second at higher temperatures due to quantum fluctuations.
One possible explanation for the single linear region is that the Bose glass contribution is immeasurably
small.  This would be consistent with the observation made above that
the strong Van der Waals potential  makes the $^4$He film nearly incompressible.

Assuming for the moment that the Bose glass contribution to the heat
capacity is too small to be measured, we turn to the linear behavior
that we do observe.  At onset this is
predicted by the scaling theory
of Fisher {\it et al.}, in which the contribution to $C$ from 
quantum fluctuations varies linearly with temperature.  We note, however,
that the linear behavior is observed over a very wide temperature range,
extending up to at least 600~mK, while the superfluid density data for both 
aerogel and Vycor indicate that the critical regime in which 
quantum fluctuations are relevant is confined to coverages with
$T_c$'s less than 200~mK. Although there is no reason why the asymptotic  
critical regions observed for two different thermodynamic quantities
must be the same, it would be rather unusual to see critical
behavior so readily in $C(T)$ when all other measurements suggest
that the critical regime is nearly inaccessible.

\subsection {Activation Model for Non-superfluid Films}

As can be seen in Fig.~ \ref{fig:alldat2}, the heat capacity of non-superfluid films
drops off very rapidly, faster than $T^3$, at low temperatures.  This behavior
is suggestive of a gap in the excitation spectrum of the film.    The simplest 
density of states that includes a gap is shown in Fig.~\ref{fig:tait}.  The model,
considered originally by Tait and Reppy,\cite {Tait}  includes two bands of
states.   The lower band comprises localized states due to the heterogeneous
adsorption potential.   Because of the strong repulsive He--He interactions,
each localized state can accommodate only a single He atom, leading to an effective
Fermi statistics in the lower band of Fig.~\ref{fig:tait}.   The upper band, which
is unoccupied at $T=0$,  consists of extended states, which are assumed to obey
Bose statistics at low densities.  This model leads to a heat capacity of the
form
\begin {equation}
C = D(\Delta /2k_BT + 2)e^{-\Delta/2k_BT}
\label {eq:activation}
\end {equation}
for $k_BT \ll \Delta$.  The prefactor $D = \sqrt{N_lN_u}\Delta k_B$, where
$N_l$ and $N_u$ are the densities of states in the lower and upper bands.
The results of fits to Eq.~\ref{eq:activation} for four non-superfluid films in
the regime $T < T_B$ 
are shown in Fig.~\ref{fig:activationfits}.    The gap $\Delta$ is shown as a function 
of coverage in the inset.   For each coverage, we find a gap of order $5T_B$.  

Our data are not precise enough to distinguish
between a true gap in the density of states and a ``soft gap'', in which
a low-energy tail exists in the band of extended states.  It is 
possible that a small number of extended states exist at energies well
below $\Delta$.  In principle, it would be possible to place a bound
on the number of such states by determining how much of the classical
entropy is {\it not} frozen out at low temperatures.  Unfortunately, this
calculation is impossible because our data do not reach a limit in 
which $C(T)$ is independent of temperature.  (This
limit could not be reached without macroscopically populating the
vapor phase, which is ignored in our model.)

There is no {\it a priori} reason for  our having chosen a {\it constant} density of states
in each of the bands shown in Fig.~\ref{fig:tait}.  As far as fitting the data
of Fig.~\ref{fig:activationfits} is concerned,  the critical element of the excitation
spectrum is the gap.    Additional information, however, comes from the high-temperature
data on both sides of the onset transition.  In both cases, the heat capacity depends linearly
on temperature in this regime.  For $n < n_c$, only the extended states contribute to the
heat capacity at high temperature, and a linear temperature dependence would follow from a 
constant density of states.   The only difference for superfluid films $(n>n_c)$  is that some extended
states are occupied at $T=0$.  For $T>T_c$ we would expect to see the same  temperature
dependence as observed for films below the critical coverage.     Based on Fig.~\ref{fig:3D}, this appears to be the case, since the wedge-shaped plateau at high-temperatures extends to both sides of the critical coverage.  We have also fitted the
high-temperature ($T>2T_B$ or $T>T_c$) data to the form $C(T) = AT + BT^2$   on
both sides of the onset transition.  As can be seen in Fig.~\ref{fig:ABfits}, the linear
coefficient $A$ increases slowly with coverage, with no apparent feature at the onset
transition, which is marked in the figure by the dashed line.  The quadratic contribution,
however, increases rapidly for coverages above onset, although it remains at least
a factor of two smaller than the linear heat capacity for $T < 400$~mK.     

The onset of superfluidity in $^4$He films adsorbed in Vycor thus corresponds to
the closing of a gap in the excitation spectrum of the film as the coverage is increased.
Although this is a Bose insulator to superfluid transition, the insulating phase is not
gapless, in disagreement with the assumptions of the Bose glass model.
Although this conclusion turns to some degree on whether or not we can resolve
a small but nonzero density of states at low temperatures, there is also a significant
philosophical distinction between the activation model we have employed here
and the Bose glass approach.    In the Bose glass model, superfluidity is destroyed at $n=n_c$
by order parameter phase fluctuations. The Bose condensate (essentially the microscopic
order parameter) is assumed to exist below onset, even though long-range order
does not.    In the simple activation model employed here, there is no order parameter
below onset, since none of the extended states in the system are occupied at $T=0$.
A similar distinction appears for the case of the superconductor-insulator transition, for which two classes of transitions have been observed.
For granular films, the order parameter is non-zero at the
superconductor-insulator transition,\cite {White}  and the superconductor-insulator
transition is driven by  fluctuations in the phase of the order parameter.  There is also evidence for a Bose
glass phase in some homogeneously disordered  films,\cite {Paalanen} but other
experiments suggest that the amplitude of the order parameter vanishes at the 
superconductor-insulator transition in this case. \cite {Valles}   At the moment, we cannot
probe the microscopic order parameter in our system, and so we cannot determine directly whether or not a condensate exists for coverages below the onset of superfluidity.

\subsection {Thin Superfluid Films}

As discussed above, the rapid change of the chemical potential $\mu$ with
coverage near $n_c$ tends to make quantum fluctuations less important
in $^4$He films than they would be if the adsorption energy
were less strongly dependent on coverage.
The large spread in energy scales due to the adsorption potential
minimizes exchange between  atoms
in the localized layer and those  in the overlying superfluid film.   We believe this
lies behind the qualitative success of the inert layer model
for thicker films, and it is reasonable to ask how close
to the onset transition one can neglect exchange with the non-superfluid
film.    

If we ignore the $^4$He atoms in the localized layer in our aerogel sample, we estimate
that the inter-particle spacing $d$ for the atoms in the superfluid
film is at least 10~\AA\ for $T_c < 80$~mK.  This estimate follows
from dividing the number of superfluid atoms by the {\it surface
area} (9.2~m$^2$)  of the substrate.  Estimates based on the total {\it volume}
occupied by the substrate (0.13~cm$^2$) give
an interparticle spacing on the order of 
30~\AA\ for a film with $T_c$ = 80~mK.\cite {PACthesis}  Both of these estimates
indicate that it is not unreasonable to look for dilute Bose-gas like
behavior, since the $s$-wave scattering length characterizing the He-He
interaction is of order 3~\AA .

A full formalism
for the crossover from a strongly interacting superfluid  to a dilute Bose gas as the 
$^4$He coverage
is reduced has been developed by Weichman and co-workers.\cite {WRSF} The starting point for this theory is the observation that thin films of $^4$He
adsorbed on Vycor show 3D XY critical behavior for coverages with transition
temperatures above 120~mK.\cite {Bishop}  Identical critical behavior
(for films with $T_c$'s down 50~mK) has been observed in $^4$He films adsorbed
in 91\%\ porosity aerogel glass.\cite {Crowell}   In both systems,  power-law behavior 
of the superfluid density is observed over at least one order of magnitude in reduced 
temperature.  Weichman {\it et al.}  derive a scaling function which
in principle can be used to test for crossover from conventional 3D XY critical
behavior to dilute Bose gas behavior as the $^4$He density decreases.
The scaling function reduces to the dilute Bose gas form  
$\rho_s(T) \sim \rho_s(0)[1-(T/T_c)^{3/2}]$ as $T \rightarrow 0$ and
to the 3D XY model form $\rho_s(T) \sim \rho_{s_0}(1-T/T_c)^\zeta$
($\zeta \approx 2/3$) as $T \rightarrow T_c$.   Physically, the scaling function
accounts for the reduction in the size of the 3D XY critical regime as $T_c$ decreases.
Unfortunately, the Bose gas and 3D XY limits are sufficiently distinct that a scaling
function capable of accounting for the crossover can be used to collapse data from
essentially any superfluid film.\cite {KW}
The two limiting forms, however, suggest
another way of testing for crossover to Bose gas-like behavior.\cite {Crooker}
The quantity $\lim_{T \rightarrow T_c} \rho_s(T)/(T_c - T)$ {\it diverges}
in the strongly interacting limit but remains finite for the dilute
Bose gas.  In Fig.~\ref{fig:dbg}, we plot $\Delta P(T)/(T_c - T)$  as a function of $T/T_c$ 
for four coverages of $^4$He adsorbed on aerogel.    The dilute Bose gas prediction
(assuming the same ratio $\Delta P(0)/T_c = 12$~psec/mK found for the four films) is
shown as a dash-dotted line.   The divergence near $T_c$ is pronounced for all of the films.

There is thus no convincing evidence that $^4$He films adsorbed in
aerogel approach Bose-gas like behavior for films with $T_c$'s
down to 20~mK.  The absence of any observable crossover may be
due to exchange with the localized layer
or simply the possibility that our films were not sufficiently
dilute.  Crooker {\it et al.} studied films with $T_c$'s down
to 7~mK in their study of the $^4$He-Vycor system and found
a progressive decrease in the 3D XY critical regime as determined from 
the same analysis used here to produce Fig.~\ref{fig:dbg}.  Unfortunately,
the superfluid signal for a given substrate volume is smaller
for aerogel than for Vycor, and the thermal time constants are much
longer in aerogel, so extending our measurements to 
lower coverages and temperatures would not be a trivial task.
This is unfortunate, since
the superfluid transitions in $^4$He-aerogel are ``sharper''
than those in $^4$He-Vycor, in the  sense that $\rho_s$  goes
to zero only 0.2-0.3~mK above $T_c$  (as determined from
power-law fits\cite {Crowell})
as opposed to 1-2~mK in Vycor.  Kotsubo and 
Williams\cite {KW} have noted that the smearing of the
transition in the Vycor case tends to mimic crossover to 
Bose gas-like behavior.  There is now convincing evidence
that the 1-2~mK ``foot'' in the Vycor data was due to macroscopic
inhomogeneities introduced by the leaching process used to produce
the porous glass.\cite {Robinson}  
A new Vycor study, with more homogeneous samples,
would probably be much more conclusive than the experiment of
Crooker {\it et al.} in testing for crossover to dilute Bose gas behavior.  Porous gold substrates,\cite {Yoon} which recent full-pore
measurements suggest are as homogeneous as Vycor, could be an even
better substrate for ultra-low temperature measurements.   

The difficulties in interpreting the $\rho_s$ measurements
for very thin superfluid films have been compounded by the absence of a
complementary study of other thermodynamic quantities in this
coverage regime.  Our heat capacity measurements for
$^4$He-Vycor begin to address this problem, although the
lowest transition temperature we were able to achieve was only 
21~mK, significantly higher than the lowest $T_c$ ($\sim 7$~mK) 
reached in the $\rho_s$ study of Crooker {\it et al.}\cite {Crooker}

In Fig~\ref{fig:21and85mK}, we show heat capacity data for two superfluid coverages,
$n=28.8$~$\mu$moles/m$^2$ ($T_c =$~85~mK) and 
$n=27.6$~$\mu$moles/m$^2$ ($T_c = $~20.7~mK).
The heat capacity is divided by temperature and is plotted as a function of the
reduced temperature $(T-T_c)/T_c$.    The heat capacity for both coverages
crosses over from a predominantly linear dependence on temperature above
$T_c$  to a  quadratic dependence for $T < T_c$.  In addition to these 
``backgrounds", there is a peak in the upper data set ($T_c =$~85~mK)
at the transition.    A peak can only barely be resolved for the film
with $T_c = 20.7$~mK.  
This trend is consistent with that observed
in earlier studies,\cite {Finotello,Murphy}  in which the magnitude of the
heat capacity peak 
for $^4$He films adsorbed in Vycor was shown to scale according to the
hyperuniversality relation,\cite {SFW}
\begin {equation}
\frac {C^\prime_s(t)}{C_s(t)} = \frac {\xi(t)^3}{\xi^\prime (t)^3},
\label {eq:tsfu1}
\end {equation}
for two systems, denoted primed and unprimed, within the same universality
class.
In Eq.~\ref{eq:tsfu1},
$t$ is the reduced temperature $(T_c - T)/T_c$  and $\xi$ is the correlation
length.    Correlation lengths for the $^4$He-Vycor system can be estimated  
from the Josephson relation,\cite {Josephson}
\begin {equation}
\xi(t) = \left (\frac {k_B T m^2}{\hbar^2} \right )\frac {1}{\rho_s(t)},
\end {equation}
and $\rho_s$ data from the studies of Bishop {\it et al.},\cite {Bishop}
and Crooker {\it et al.}\cite {Crooker,Crookerthesis}  Based
on these studies, we estimate that the correlation length for a given
reduced temperature will be approximately twice as large for the film with 
$T_c = 20.7$~mK than for the film with $T_c = 85$~mK.\cite {Crookerthesis}
According to the hyperuniversality relation Eq.~\ref{eq:tsfu1}, the heat
capacity peak for the thinner film should therefore be eight times smaller than
for the thicker film, which is consistent with Fig.~\ref{fig:21and85mK}.  

A rigorous test of Eq.~\ref{eq:tsfu1}
for $^4$He  films in Vycor would require fitting each of the heat capacity data sets to power
laws near $T_c$.   This is impractical at the moment due to both the noise in the data
and the ambiguities in subtracting a non-singular background.  Although the noise
could be reduced in a better experiment, the interpretation of the background
heat capacity poses a more fundamental problem.    In Fig.~\ref{fig:21and85mK},
the data for $T_c = 20.7$~mK show a cusp at $T_c$, where $C(T)$ crosses
over abruptly from quadratic to linear dependence on temperature.   This cusp
may be  the remnant of the fluctuation peak seen for thicker films, in which
case it should round out as the superfluid coverage is further reduced.  Another
possibility is that the heat capacity is singular even in the absence of order parameter
fluctuations.  Although this type of behavior would emerge in any mean-field treatment,
for an interacting $^4$He film we would expect a jump discontinuity
in $C(T)$ in addition to a change in slope.  (The ideal Bose gas should show only a simple cusp at $T_c$,
but of a different form than that seen in Fig.~\ref{fig:21and85mK}.)

\section {Conclusions}

The superfluid density and heat capacity measurements presented in this paper indicate
that a $^4$He film adsorbed on a disordered substrate undergoes a transition from a 
Bose insulator to a superfluid at the critical coverage $n_c$.    The gap in the insulating
phase decreases to zero as the onset coverage is approached from below.
Our data show no evidence
of a Bose glass phase.   In considering thin superfluid films, the rapid change of the
chemical potential with coverage near $n_c$  appears to minimize the role of quantum
fluctuations, so that the ordinary superfluid transition at $T_c$ remains the dominant
critical point.   Evidence of critical fluctuations near $T_c$ is seen even for the
thinnest superfluid films studied in these experiments.

Although some of the questions raised in this work could  be addressed in experiments
at even lower temperatures and smaller reduced coverages, the use of different substrates
may be more rewarding.   We have argued here that the strong substrate potential  in porous glasses reduces the importance of quantum fluctuations near the critical coverage.
Weakly binding alkali metal substrates\cite {alkali} may ameliorate this effect.   In principle, the use of
ordered substrates would allow a test of the clean limit of the various dirty boson models,
provided that the adsorption potential is sufficiently weak.  Such a system might  be achieved
by pre-plating a clean high-surface area substrate, such as exfoliated basal-plane graphite, with 
molecular hydrogen.  Finally, even on a strong-binding ordered substrate such as graphite,
the adsorption potential changes only weakly within each monolayer, so that quantum effects should be more readily observable.    The phase-separated superfluids which exist within the second and third monolayers of $^4$He on graphite may therefore be reasonable analogs of granular superconductors.\cite {Huberman}

\acknowledgments

We thank Mary Lanzerotti for assistance with the superfluid density measurements.  PAC
acknowledges financial support from AT\&T Bell Laboratories through its PhD scholarship program.  
This work was supported
by the National Science Foundation under Grant Nos. DMR-89-21733 and DMR-93-03855 and
by the MRL program of the National Science Foundation under Award
No. DMR-91-21654.


\begin{figure}
\caption {Generic phase diagram in the density-temperature plane 
for a $^4$He film adsorbed on a disordered substrate.}
\label {fig:generic}
\end {figure}

\begin {figure}
\caption {Phase diagram for the Bose-Hubbard model in the limit of strong disorder in
the plane of chemical potential $\mu$ and exchange strength  $J$.  Both coordinates are
scaled by the strength $V$ of the repulsive interaction.  The transition from the Bose-glass
(BG) to the superfluid (SF) phase occurs at the critical chemical potential $\mu_c$ as the
system travels along a path of nearly constant $J$, as would be appropriate for the onset
transition in a $^4$He film.}
\label{fig:glass}
\end {figure}

\begin {figure}
\caption {The resonant period (open circles) and amplitude (closed circles)  of the torsional oscillator
as a function of temperature during a drift measurement at a coverage of 35.7~$\mu$moles/m$^2$.  The period has been offset by -3.992359~msec.}
\label {fig:drift}
\end {figure}

\begin{figure}
\caption {A cross-section of the $^4$He-Vycor calorimeter.  All pieces are made of silver except as indicated.}
\label {fig:calorimeter}
\end {figure}

\begin {figure}
\caption {The CMN magnetization thermometer.}
\label {fig:thermometer}
\end {figure}

\begin {figure}
\caption {The response of the calorimeter following heat pulses at 102.1~mK (upper panel) and 18.6~mK (lower panel).  The solid curves show exponential fits to the decay curve at
long time scales.  Note the deviations at short time scales in the 18.6~mK data.}
\label {fig:decay}
\end {figure}

\begin {figure}
\caption {Heat capacity, divided by temperature, for a coverage of 27.6~$\mu$moles/m$^2$ of $^4$He on Vycor.  The cusp at 20~mK is a signature of the
superfluid transition.}
\label {fig:cpexample}
\end {figure}

\begin {figure}
\caption {The superfluid period shift $\Delta P$ as a function of 
temperature for twelve coverages of $^4$He adsorbed on 91\%\ porosity aerogel
glass.}
\label {fig:alldat}
\end {figure}

\begin {figure}
\caption {The superfluid transition temperature $T_c$ and the 
period shift at $T=0$, $\Delta P(0)$, as a function of the the $^4$He coverage
for films adsorbed in aerogel.  The solid curves are fits to the power laws of
Eqs.~\protect\ref{eq:tcpowerlaw} and~\protect\ref{eq:dp0powerlaw}.}
\label {fig:deltap0}
\end {figure}

\begin {figure}
\caption {The superfluid transition temperature $T_c$ as a function of the period shift
at $T=0$, $\Delta P(0)$.   The solid curve is a fit to
Eq.~\protect\ref{eq:tcp0powerlaw}.}
\label{fig:tcvsdeltap0}
\end {figure}

\begin {figure}
\caption {$\Delta P(0)$ as a function of the $^4$He coverage for films with 
transition temperatures up to 1~K.  The solid curve  is a linear fit to the data for films
with $T_c$'s greater than 150~mK.}
\label {fig:dp0all}
\end {figure}

\begin {figure}
\caption {The superfluid transition temperature $T_c$ as a function of 
the areal superfluid density at $T=0$, $\rho_s(0)$  for films with transition temperatures up to 1~K. 
 The solid curve is a linear fit to the data for films with $T_c <  300$~mK.  The dashed curve is the KTN line 
$\rho_s(T_c^-)/T_c = 8.73$~$\mu$moles~m$^{-2}$~K$^{-1}$.} \label {fig:tcall}
\end {figure}

\begin {figure}
\caption {The heat capacity, divided by temperature, is 
shown as  a function of temperature for 10 coverages of $^4$He adsorbed on Vycor
glass.    The coverages in $\mu$moles/m$^2$ are given in the legend. 
Non-superfluid coverages are shown in the upper panel and superfluid coverages
are shown in the lower panel.}
\label {fig:alldat2} 
\end {figure}

\begin {figure}
\caption {The superfluid transition temperature $T_c$ (circles) and the crossover temperature $T_B$ (triangles) defined in the text are shown as  a  function of coverage
for $^4$He on Vycor.  The onset coverage $n_c$ determined by extrapolating the $T_c$ data is indicated with the arrow.}
\label {fig:tbtc}
\end {figure}

\begin {figure}
\caption {A three-dimensional representation of the $C/T$ data for $^4$He in Vycor for
$T < 200$~mK over the coverage range of our experiment.}
\label {fig:3D}
\end {figure}

\begin {figure}
\caption {The period shift data for coverages of $^4$He in aerogel with the period shift 
scaled by $\Delta P(0)$ and the temperature scaled by $T_c$.  Open triangles correspond to a coverage of 52.7~$\mu$moles/m$^2$, for which $T_c = 955$~mK.  These data were obtained during a separate run.    All closed circles are for coverages with $T_c < 200$~mK.  The lowest
coverage in this figure is  35.2~$\mu$moles/m$^2$, for which $T_c = 26$~mK.}
\label {fig:bgscale}
\end {figure}

\begin {figure}
\caption {A model single-particle density of states for  
non-superfluid $^4$He films adsorbed in Vycor.\protect\cite {Tait}  The states
below $E_0$ are localized, and the states above $E_0+\Delta$ are extended.}
\label {fig:tait}
\end {figure}

\begin {figure}
\caption {Low temperature heat capacity data for non-superfluid
coverages of $^4$He in Vycor.  The solid curves are fits to the activation model
discussed in the text.  The coverages  in $\mu$moles/m$^2$ are indicated in the
legend.  The inset shows the gap (see Eq.~\protect\ref{eq:activation}) as a function of the
coverage.}
 \label {fig:activationfits}
 \end {figure}

\begin {figure}
\caption {The coefficients $A$  (circles) and $B$ (triangles) in fits of the high-temperature ($T > T_c$ or $T>2T_B$) heat capacity data to the form $C(T) = AT + BT^2$.  The arrow
indicates the onset coverage $n_c$.}
\label {fig:ABfits}
\end {figure}

\begin {figure}
\caption {Period shift data for four coverages, scaled as described in the text, compared with the predictions for a dilute Bose gas.  The Bose gas prediction is shown as  a dash-dotted line for the case $\Delta P(0)/T_c = 12$~psec/mK.  The transition temperatures
for each of the data sets are indicated in the legend.}
\label {fig:dbg}
\end {figure}

\begin {figure}
\caption {The heat capacity divided by temperature shown as a function of the reduced
temperature $(T - T_c)/T_c$ for coverages of 28.8~$\mu$moles/m$^2$ ($T_c = 85.0$~mK)
and 27.6~$\mu$moles/m$^2$ ($T_c = 20.7$~mK).}
\label {fig:21and85mK}
\end {figure}

{\narrowtext

\begin {table}
\caption {Summary of the asymptotic forms for the superfluid density
at T=0, $\rho_s(0)$, the superfluid transition temperature $T_c$ ,
and the heat capacity $C(T)$ in the limits $n\rightarrow n_c$ and
$T\rightarrow 0$.  The values given are the relevant exponents for
the Bose glass and inert layer models in two and three dimensions.}
\label{tab:scaling}
\begin {tabular} {lccccc}
Asymptotic Form & Exponent & \multicolumn{2}{c}{Bose Glass} & 
\multicolumn {2}{c}{Inert Layer} \\\cline {3-6}
 & &d=2&d=3&d=2&d=3\\
\tableline
$\rho_s(0) \sim (n-n_c)^\zeta$ & $\zeta$ & $\geq 2$ & $\geq 8/3$ & 1 & 1 \\
$T_c \sim (n-n_c)^w$ & $w$ & $\geq 2$ & $\geq 2$ & 1 & 1 \\
$T_c \sim [\rho_s(0)]^x$ & $x$ & 1 & 3/4 & 1 & 2/3 \\
$C \sim T^{\theta_-} (n < n_c)$ & $\theta_-$ & 1 & 1 & -- & -- \\
$C \sim  T^{\theta_0} (n = n_c)$ & $\theta_0$ & 1 & 1 & -- & -- \\
$C \sim T^{\theta_+} (n > n_c)$ & $\theta_+$ & 2 & 3 & 1 & 3/2 \\
\end {tabular}
\end {table}

\begin {table}
\caption {Torsional oscillator parameters for the $^4$He-aerogel measurements.
Additional details are given in Ref.~\protect\onlinecite{Crowell}.
The substrate velocity is estimated from 
the cell geometry and the various amplifier gains.}
\label {tab:cell}
\begin {tabular}{cc}
	& \\ 
\tableline
Frequency  (Hz)  & 250 \\[4pt]
 
\begin {tabular} {c} 
Moment of Inertia \\
 (g$\cdot$cm$^2$) \\
 \end {tabular}
 & 0.34 \\ [8pt]

Cell Volume (cm$^3$)  & 0.13$\pm$0.01 \\[4pt]
 
\begin {tabular} {c}
N$_2$ BET Surface \\ 
Area (m$^2$)  \\
\end {tabular} 
& 9.2$\pm$0.9 \\[8pt]

\begin {tabular} {c}
Mass Sensitivity \\
(nsec/$\mu$gram) \\
\end {tabular} 
& 0.211  \\ [8pt]

\begin {tabular} {c}
Superfluid Mass \\
Sensitivity (nsec/$\mu$gram)\\
\end {tabular}
& 0.022\tablenotemark[1] \\ [8pt]

$Q$ at 10~mK &  $1.6\times 10^6$ \\ [4pt]

\begin {tabular} {c}
Maximum Substrate \\ 
Velocity at 10~mK (cm/sec) \\
\end {tabular}
& $2\times 10^{-2}$\\ 
\end{tabular}
\tablenotetext[1]{Limiting value at high coverages.  See Ref.~\protect\onlinecite{Crowell}.}
\end {table}

\begin {table}
\caption {Experimental values of the exponents $\zeta$, $x$,  
and $w$  for aerogel and Vycor.  The Vycor exponents are based on fits to the
data of Crooker.\protect\cite{Crookerthesis}}
\label {tab:exponents2}
\begin {tabular} {ccc}
Exponent & Aerogel & Vycor \\
\tableline
$\zeta$ & $1.64\pm0.04$ & $2.5\pm0.3$ \\
$x$ & $0.95\pm0.02$ & $0.85 \pm 0.02$ \\
$w$ & $1.59 \pm 0.06$ & $1.76 \pm 0.01$ \\
\end {tabular}
\end {table}
}
\end {document}
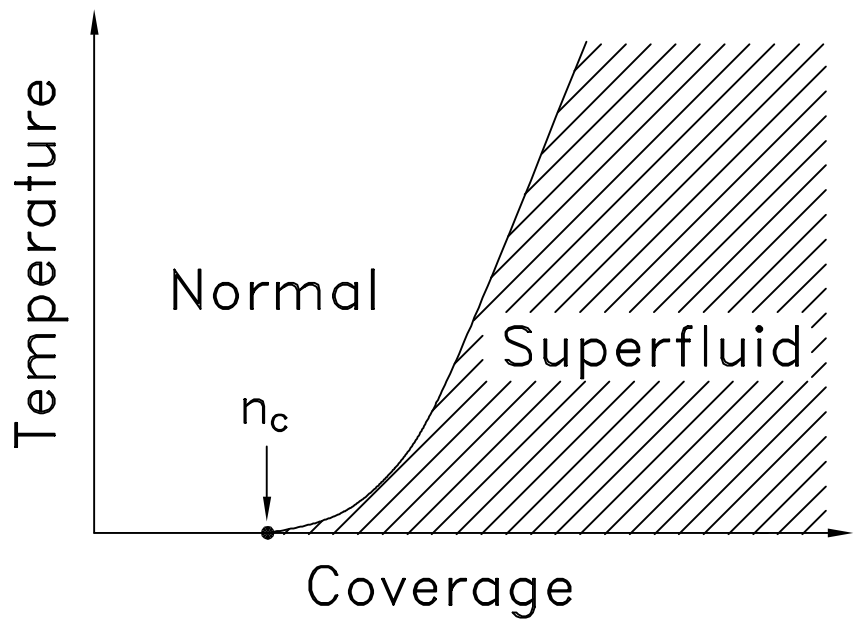
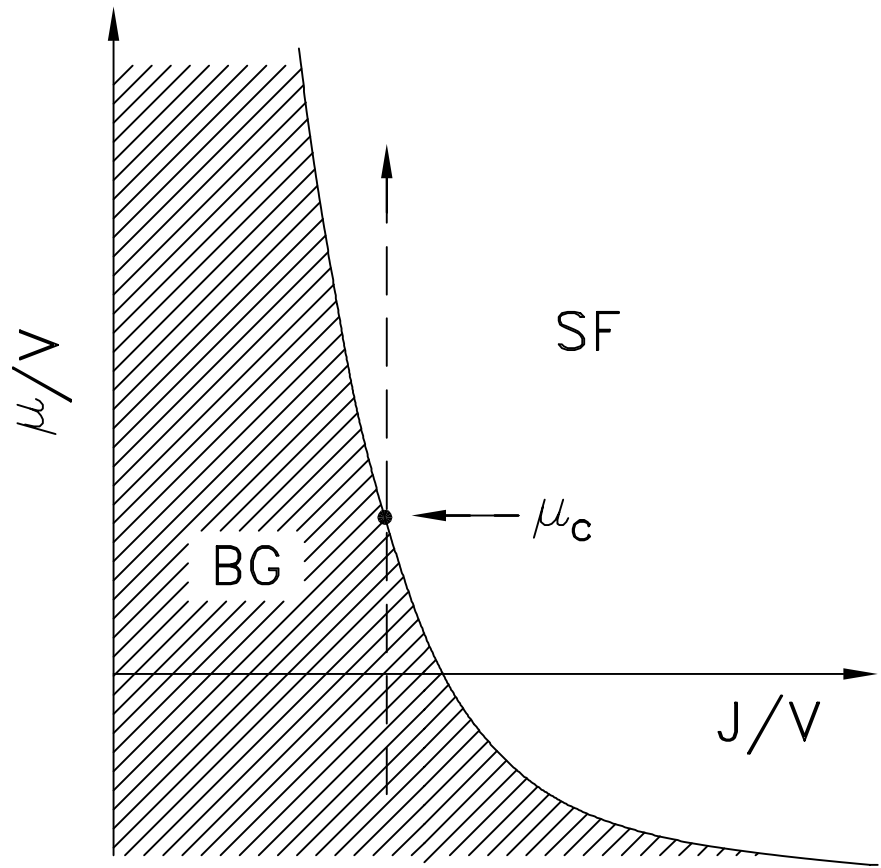
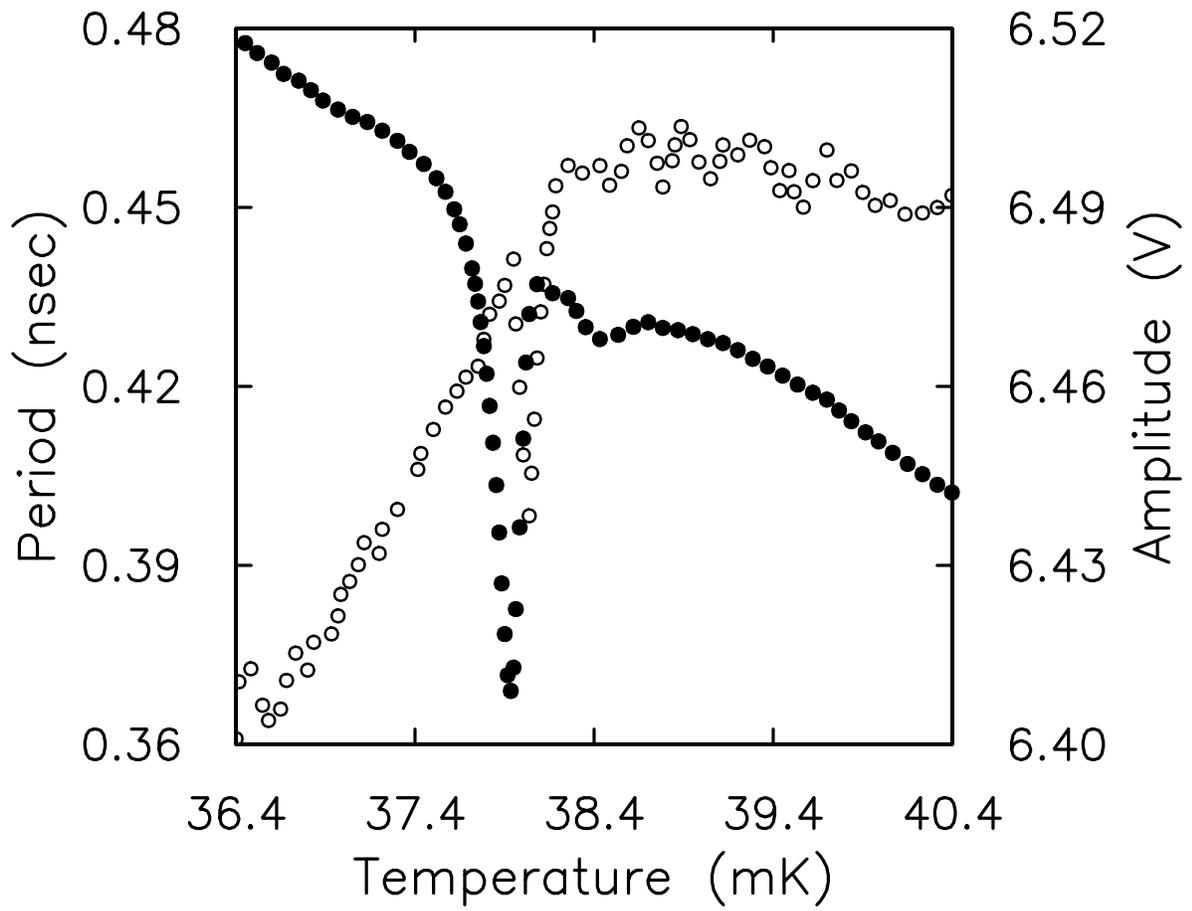
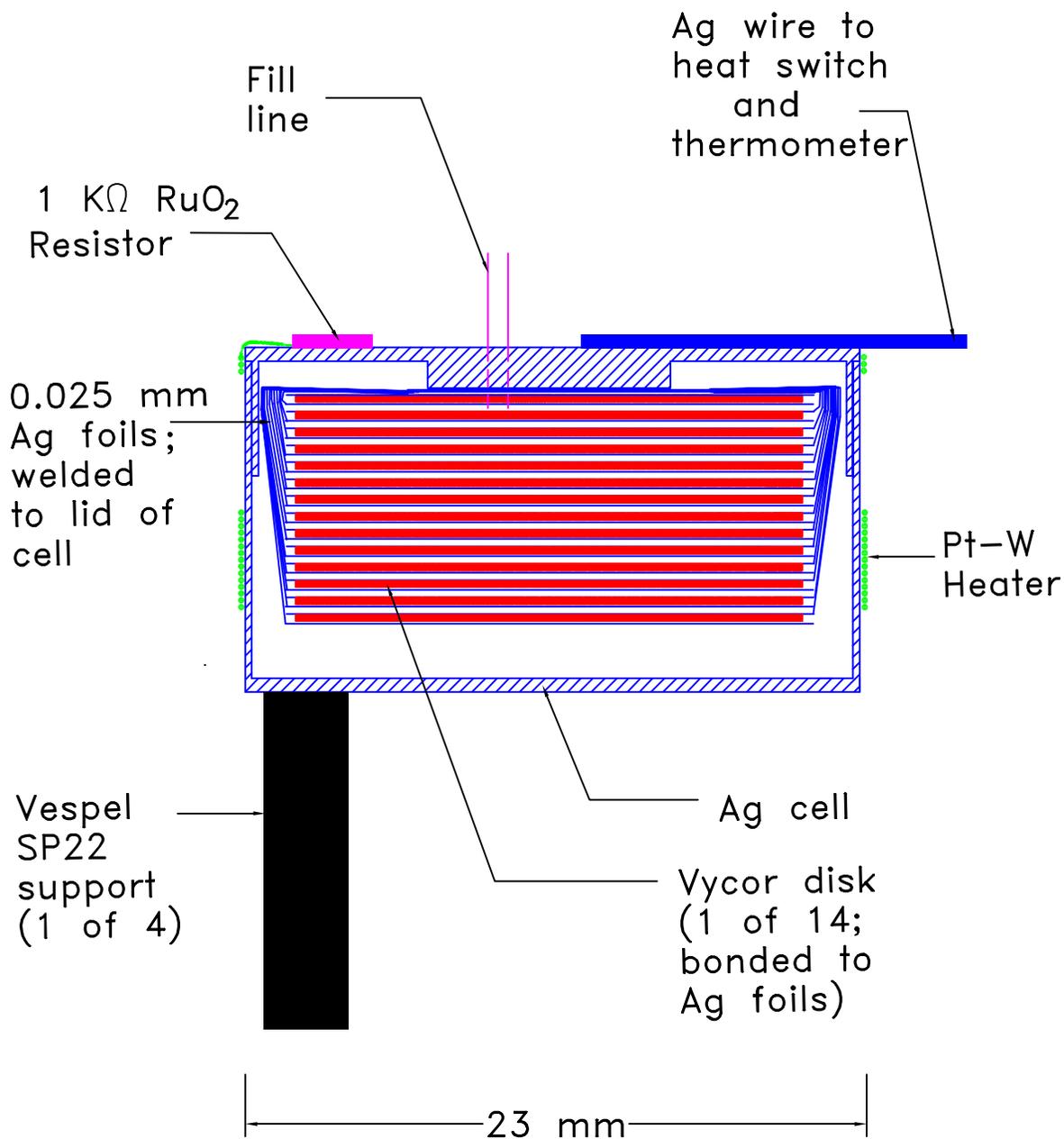
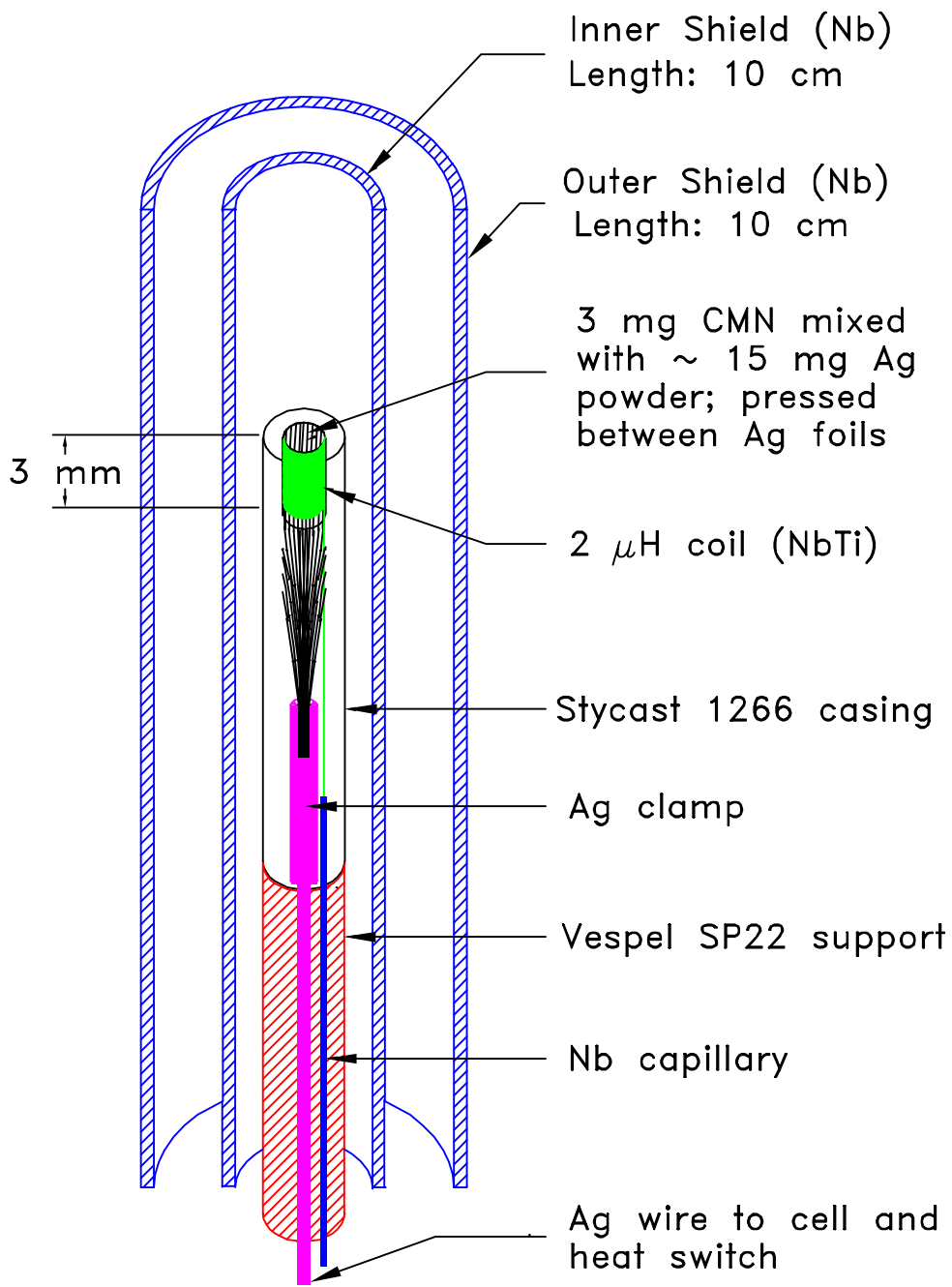
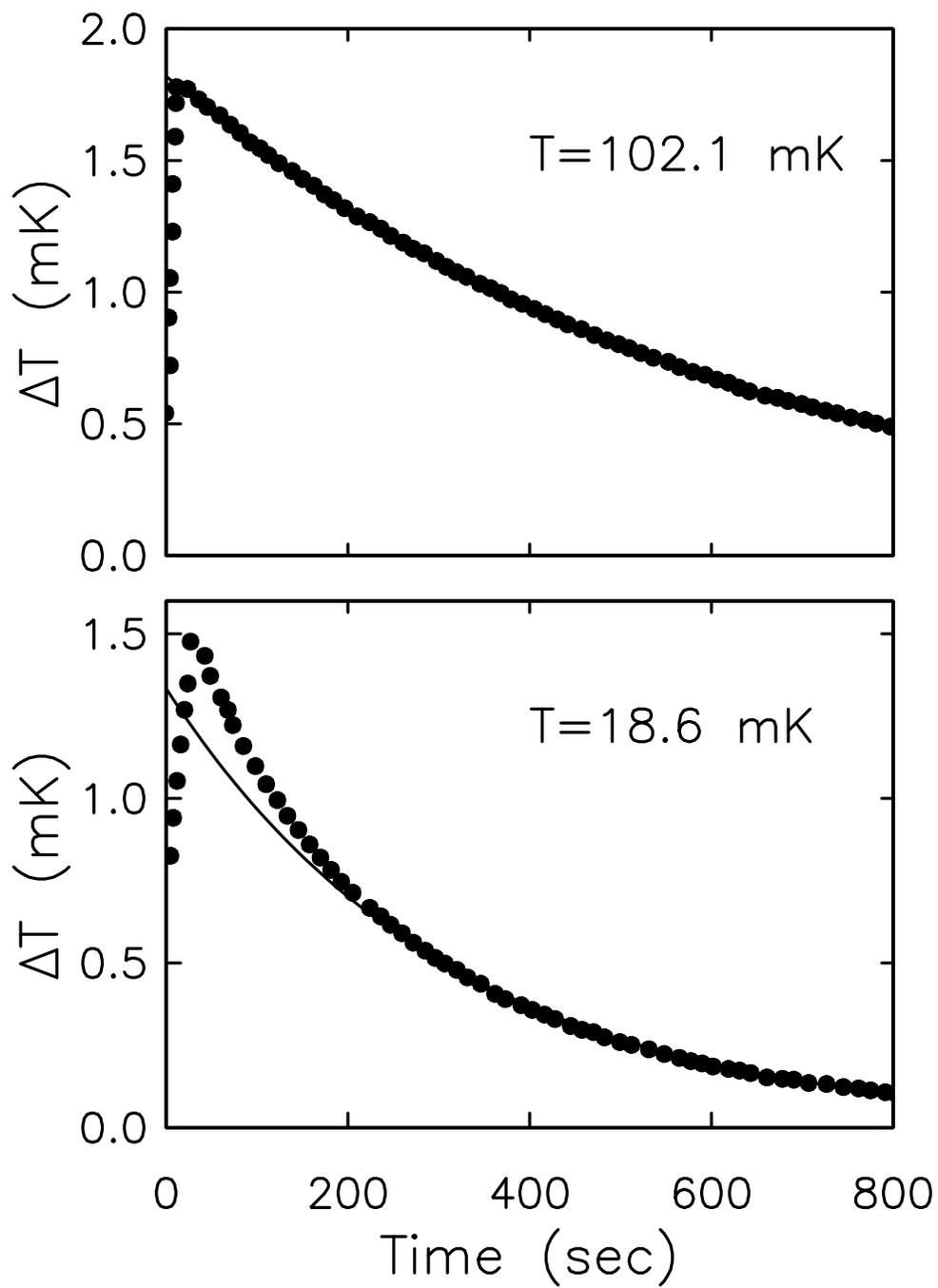
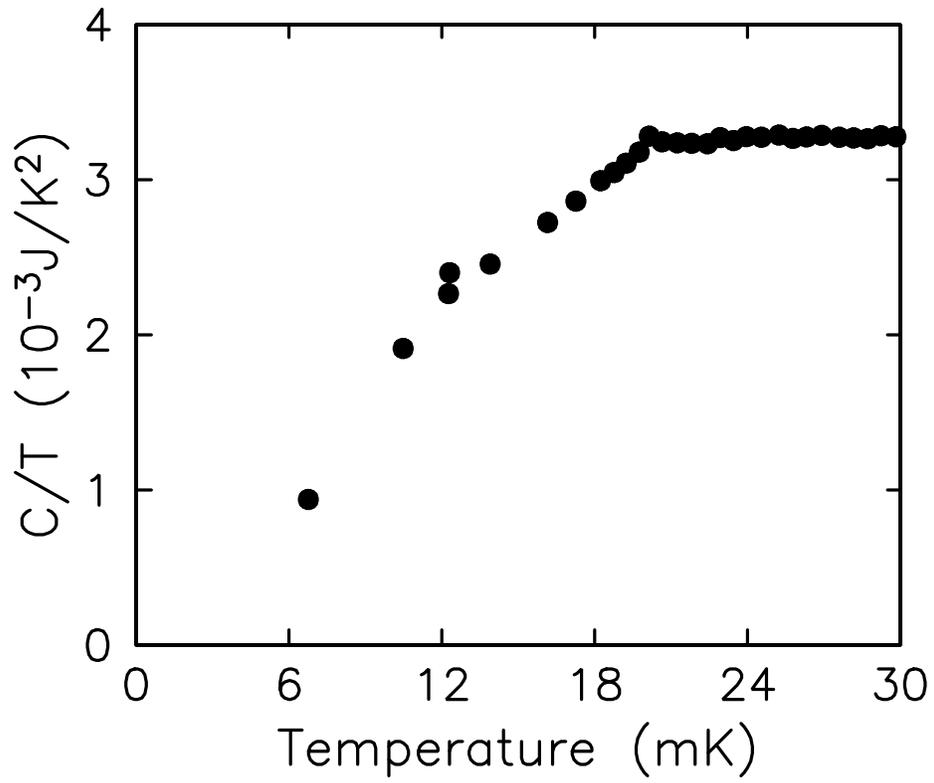
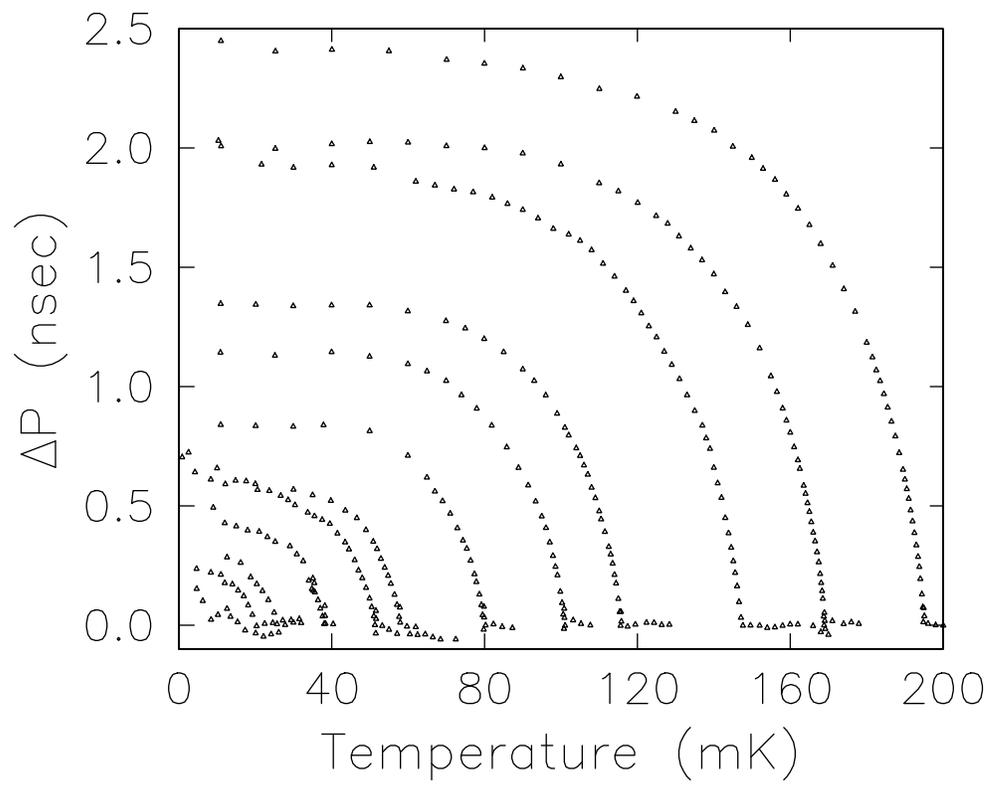
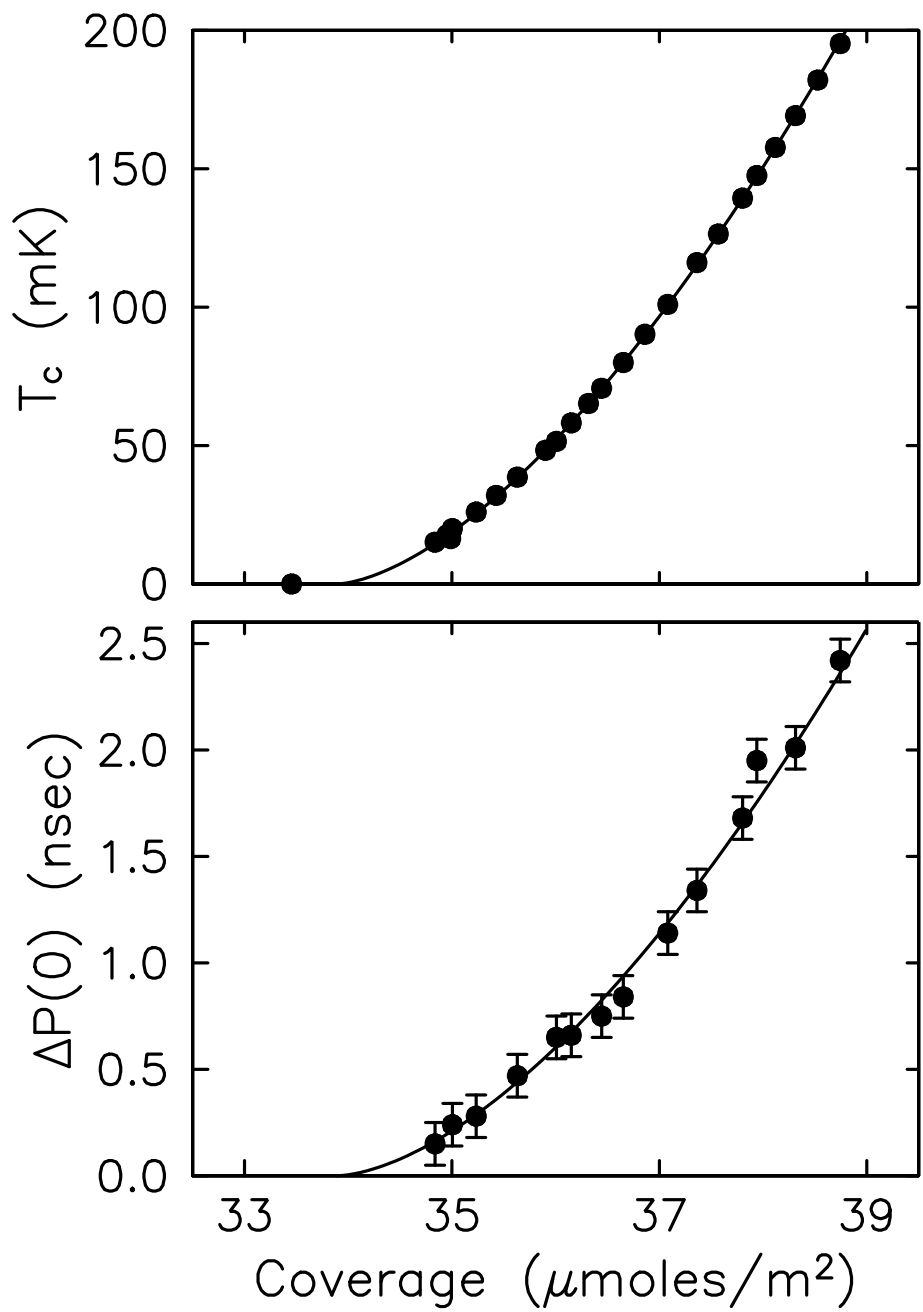
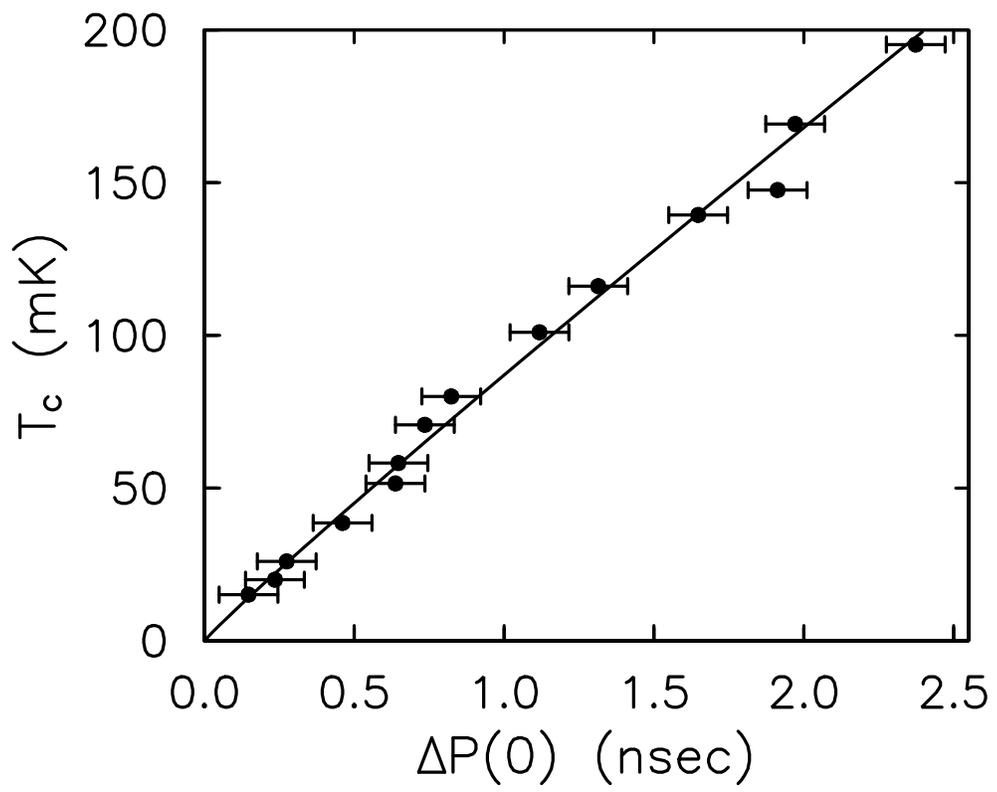
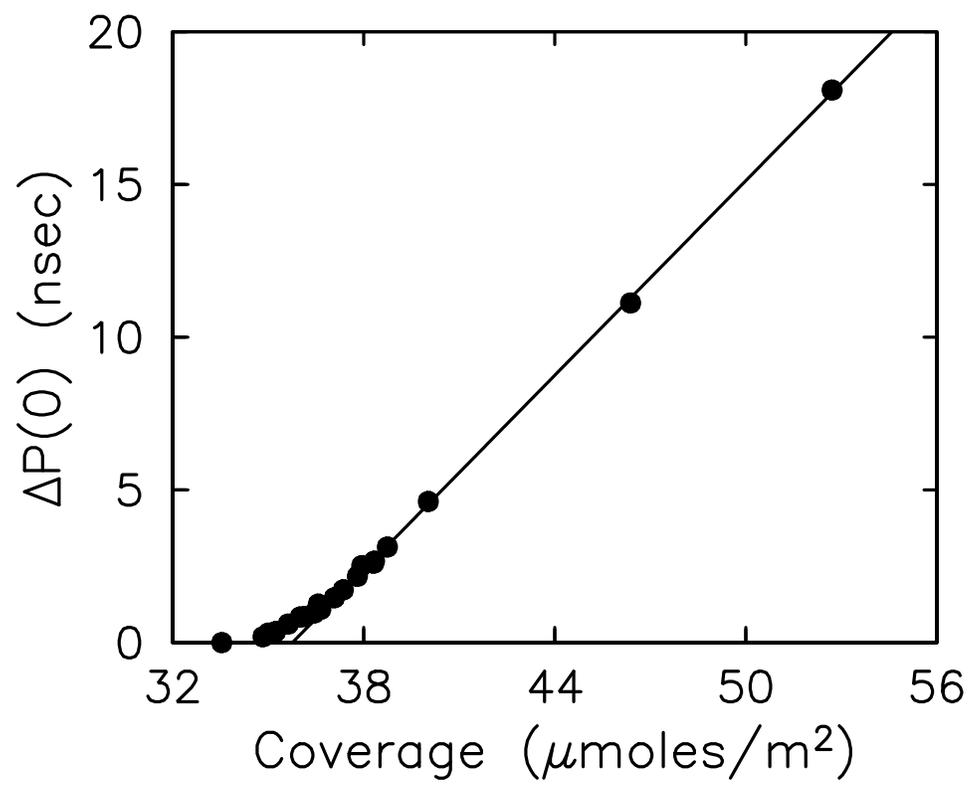
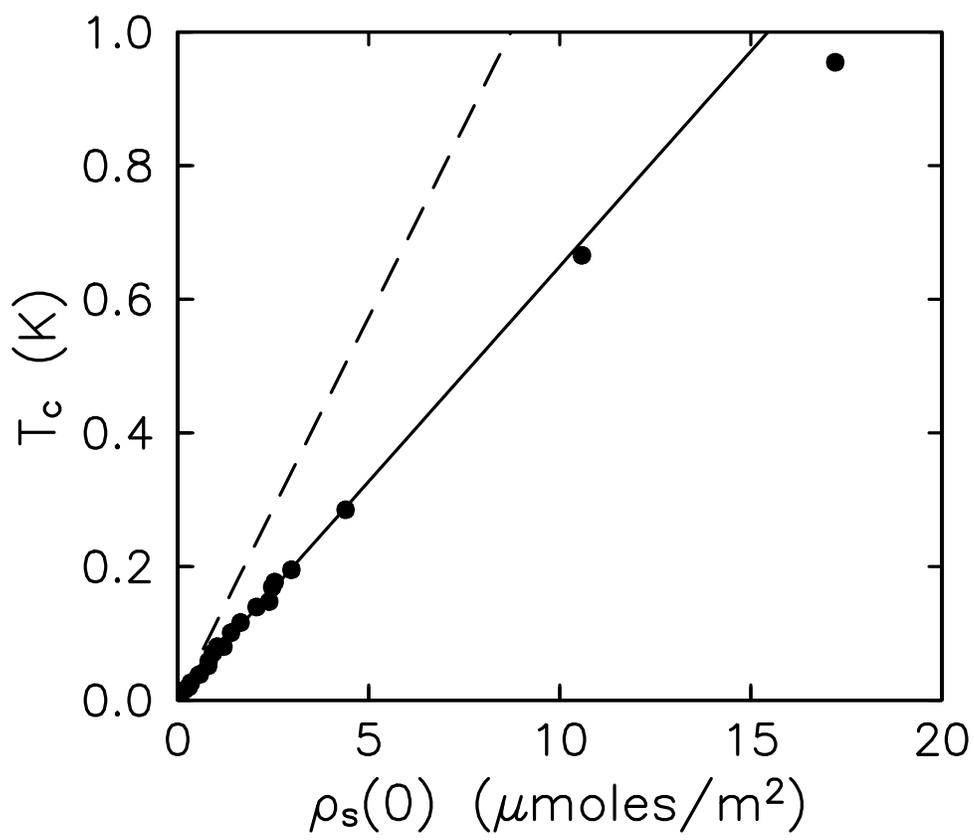
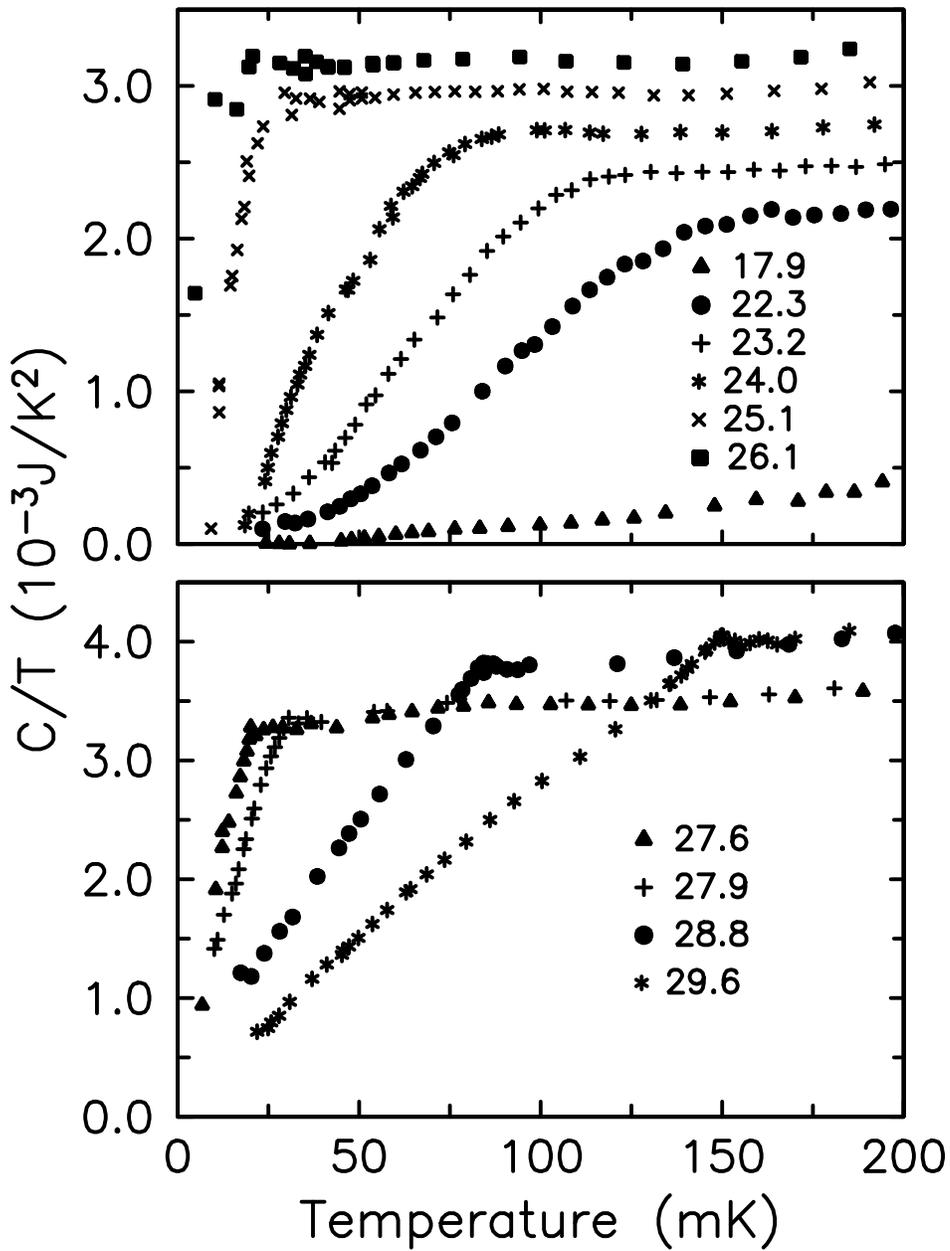
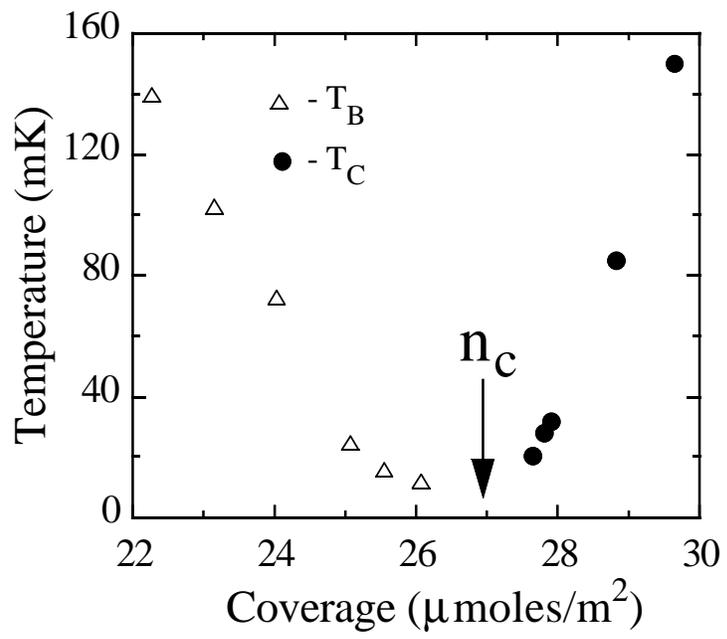
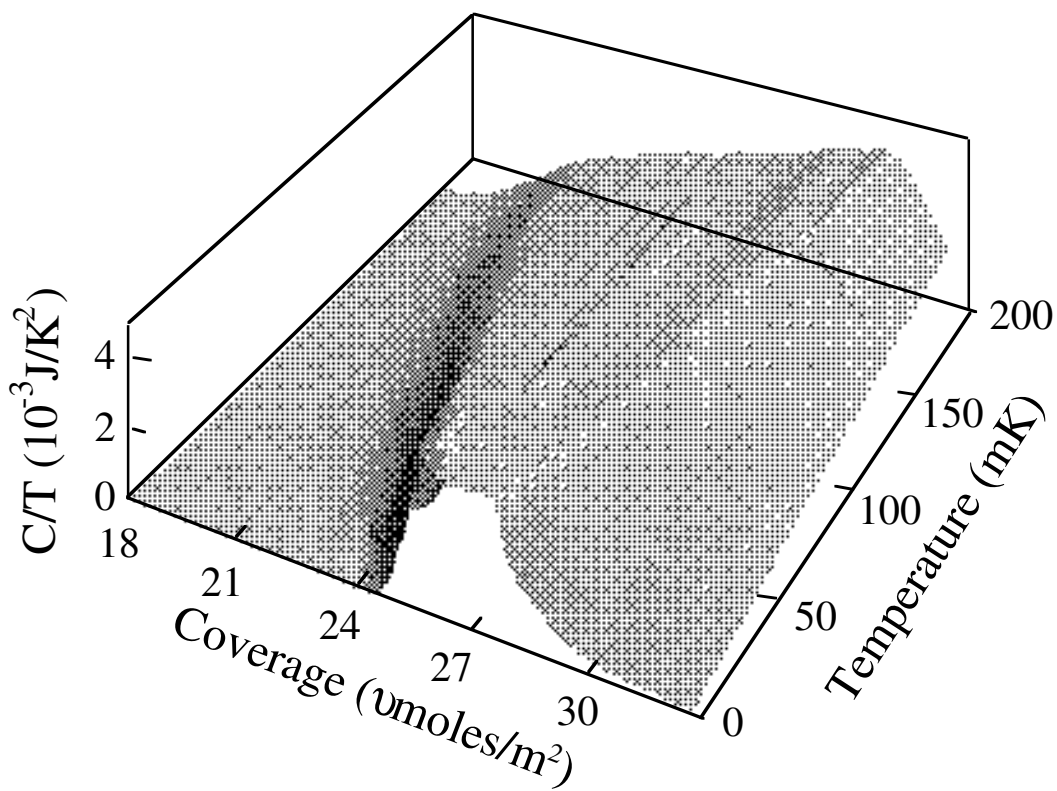
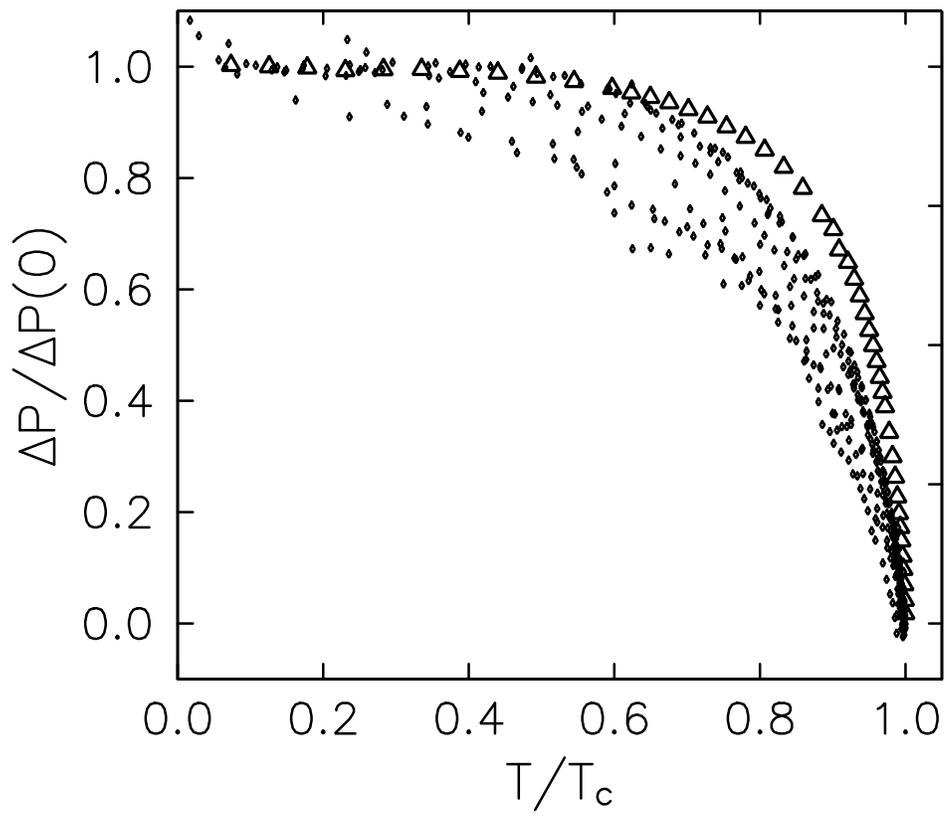
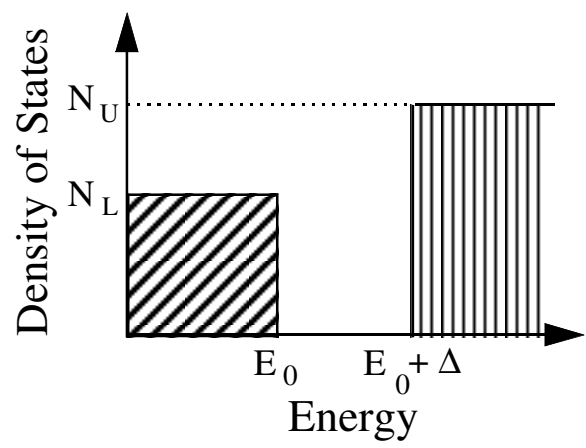
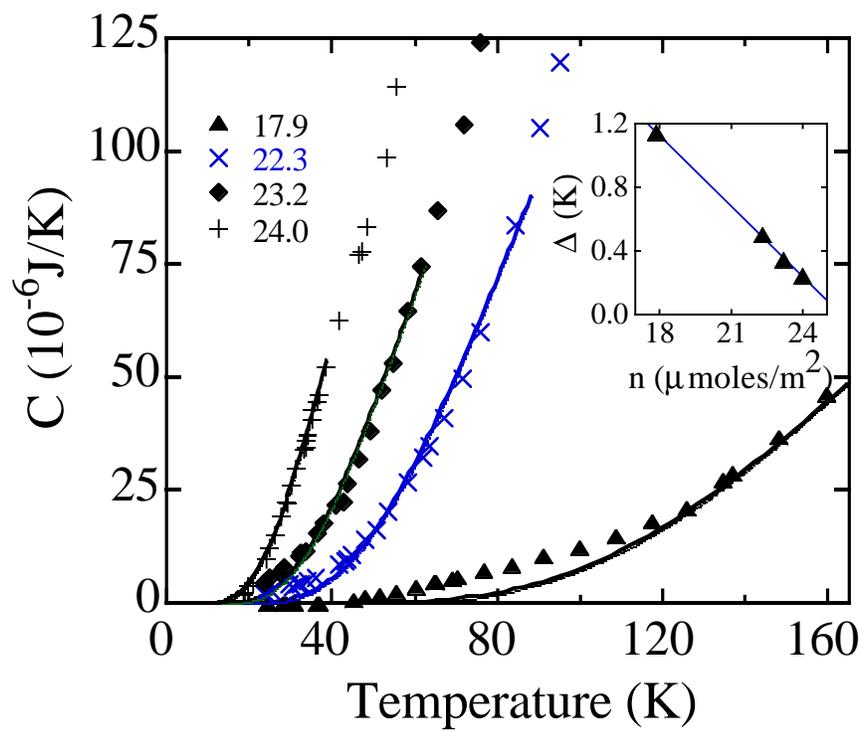
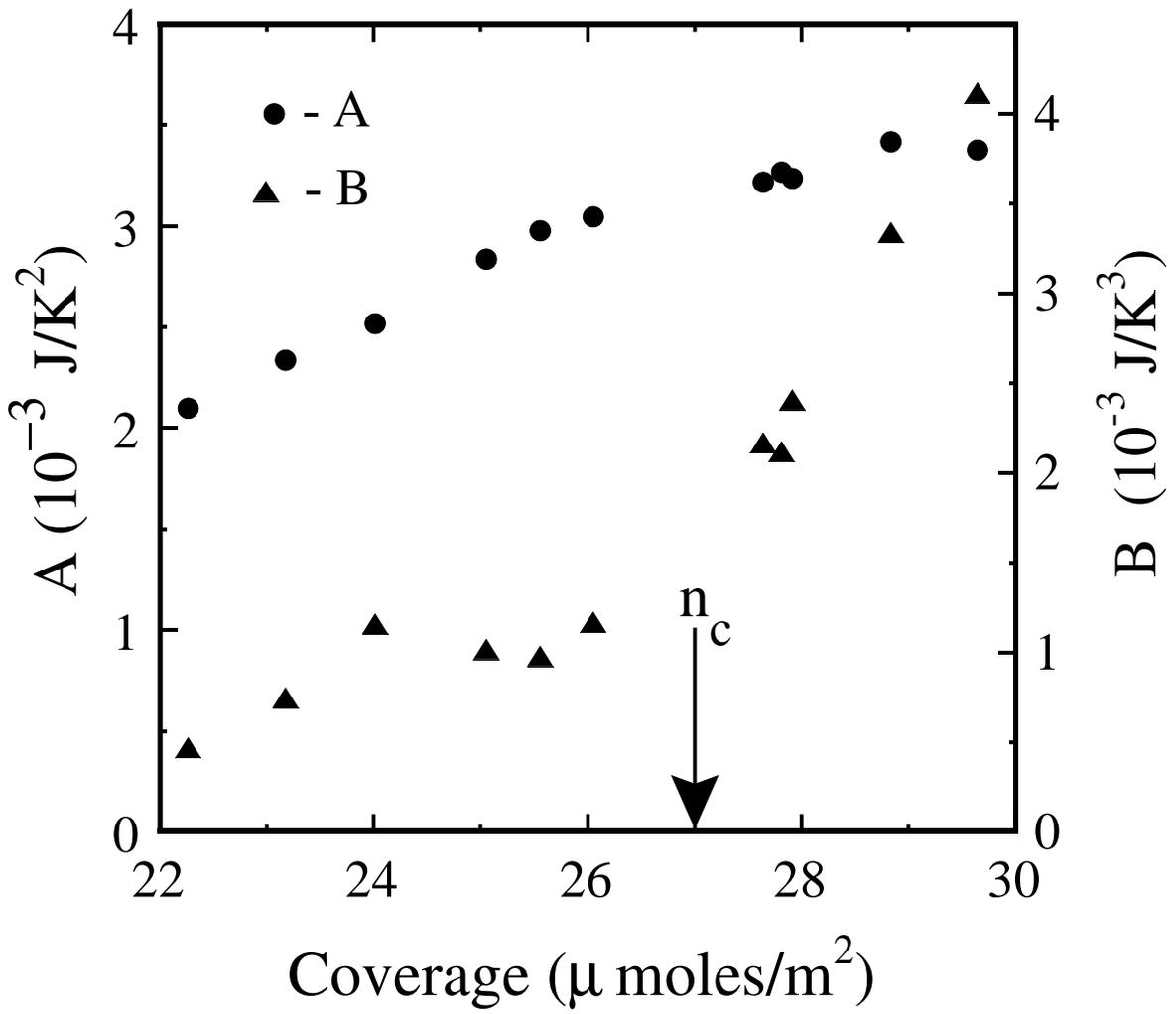
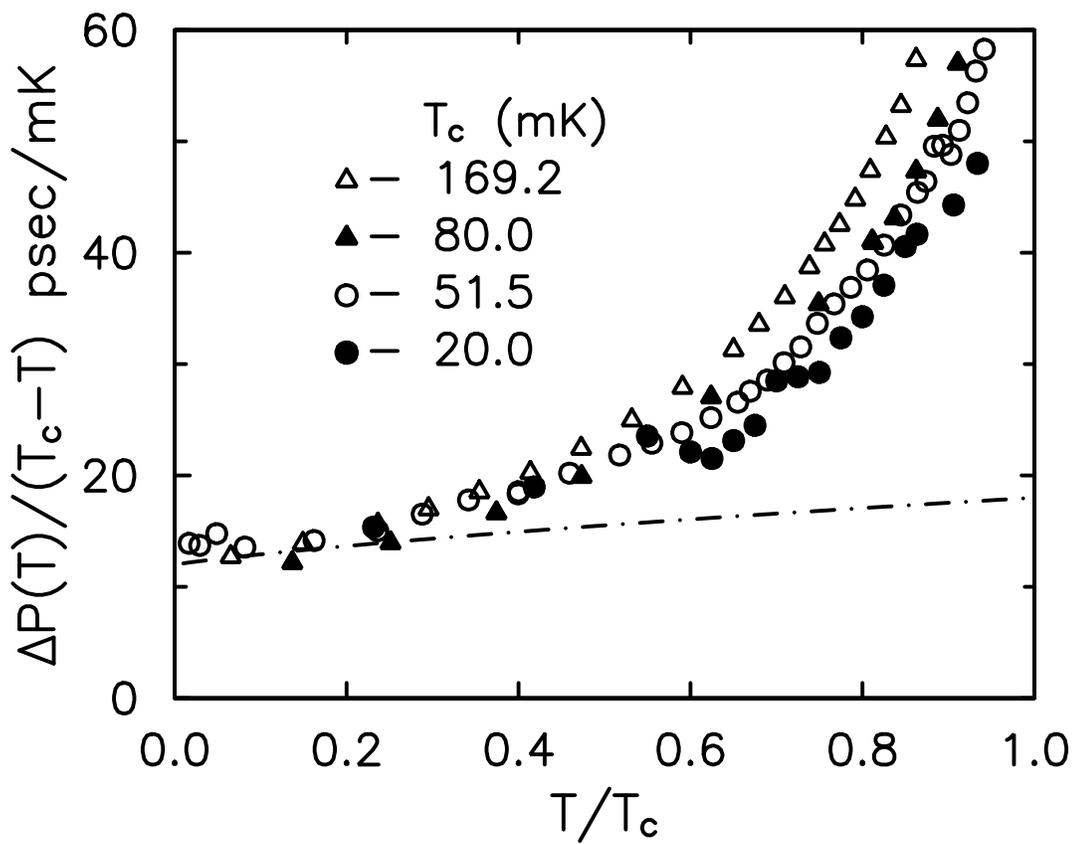
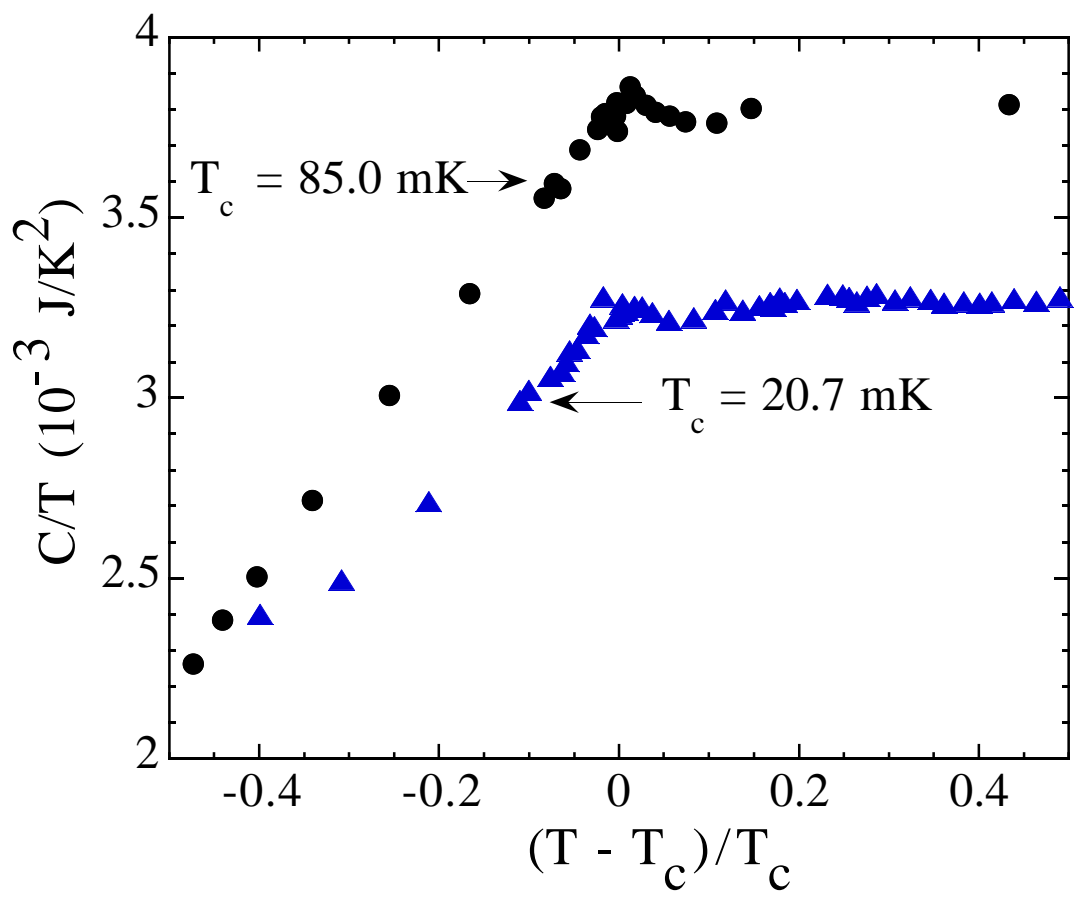
